\documentclass[usenatbib]{mnras}

\usepackage{graphicx}	
\usepackage{amsmath}	
\usepackage{amssymb}	
\usepackage{xcolor}
\usepackage{soul}
\usepackage{threeparttable}
\usepackage[normalem]{ulem}
\def\music{\texttt{MUSIC}}

\def\ahf{\texttt{AHF}}

\def\fire{\texttt{FIRE}}
\defcitealias{Hopkins:2018}{H18}
\defcitealias{Wolf:2010}{2010}

\newcommand{\rh}{\mathrm{r_{1/2}}}
\newcommand{\vh}{\mathrm{V_{1/2}}}
\newcommand{\mh}{\mathrm{M_{1/2}}}
\newcommand{\lcdm}{$\Lambda$CDM}
\newcommand{\msun}{{\rm M}_{\odot}}
\newcommand{\mhalo}{M_{\rm halo}}

\newcommand{\mvir}{M_{\rm vir}}

\newcommand{\mstar}{M_{\star}}
\newcommand{\vmax}{V_{\rm max}}
\newcommand{\vcirc}{V_{\rm circ}}

\newcommand{\hopkins}{\citetalias{Hopkins:2018}}

\newcommand{\rpower}{r_{\rm power}}

\usepackage{etoolbox}
\makeatletter
\patchcmd\@combinedblfloats{\box\@outputbox}{\unvbox\@outputbox}{}{%
  \errmessage{\noexpand\@combinedblfloats could not be patched}%
}%
\makeatother

\usepackage[T1]{fontenc}
\usepackage{ae,aecompl}

\usepackage{txfonts}

\title[Dwarf Galaxies in CDM, WDM, and SIDM]{Dwarf Galaxies in CDM, WDM, and SIDM: Disentangling Baryons and Dark Matter Physics} 
\author[A. Fitts et al.]{Alex Fitts$^1$\thanks{\href{mailto:fitts.alex@gmail.com}{fitts.alex@gmail.com}}, 
Michael Boylan-Kolchin$^1$\thanks{\href{mailto:mbk@astro.as.utexas.edu}{mbk@astro.as.utexas.edu}}, Brandon Bozek$^{1}$, James S. Bullock$^2$,\newauthor
Andrew Graus$^1$, Victor Robles$^2$, Philip F. Hopkins$^3$, Kareem El-Badry$^4$, \newauthor Shea Garrison-Kimmel$^3$, Claude-Andr\'e Faucher-Gigu\`ere$^5$, Andrew Wetzel$^6$\newauthor and Du\v{s}an Kere\v{s}$^7$\\
$^1$Department of Astronomy, The University of Texas at Austin, 2515 Speedway, Stop C1400, Austin, Texas 78712-1205, USA\\
$^2$Department of Physics and Astronomy, Center for Cosmology, 4129 Reines Hall, University of California Irvine, CA 92697, USA\\
$^3$TAPIR, California Institute of Technology, Pasadena, CA, USA\\
$^{4}$Department of Astronomy, 501 Campbell Hall, University of California, Berkeley, CA, 94720, USA\\
$^5$Department of Physics and Astronomy and CIERA, Northwestern University, Evanston, IL, USA\\
$^6$Department of Physics, University of California, Davis, CA, USA\\
$^7$Department of Physics, Center for Astrophysics and Space Sciences, University of California, San Diego, La Jolla, CA, USA\\
}
\date{\today}
\vspace{-0.5cm}
\pubyear{2018}

\begin{document}
\setstcolor{red}
\label{firstpage}
\pagerange{\pageref{firstpage}--\pageref{lastpage}}
\maketitle

\begin{abstract}
We present a suite of \fire-2~cosmological zoom-in simulations of isolated field dwarf galaxies, all with masses of $M_{\rm halo} \approx 10^{10}\,\msun$ at $z=0$, across a range of dark matter models. For the first time, we compare how both self-interacting dark matter (SIDM) and/or warm dark matter (WDM) models affect the assembly histories as well as the central density structure in fully hydrodynamical simulations of dwarfs. Dwarfs with smaller stellar half-mass radii ($\rh<500$ pc) have lower $\sigma_\star/\vmax$ ratios, reinforcing the idea that smaller dwarfs may reside in halos that are more massive than is naively expected. The majority of dwarfs simulated with self-interactions actually experience contraction of their inner density profiles with the addition of baryons relative to the cores produced in dark-matter-only runs, though the simulated dwarfs are always less centrally dense than in \lcdm. The $\vh-\rh$ relation across all simulations is generally consistent with observations of Local Field dwarfs, though compact objects such as Tucana provide a unique challenge. Overall, the inclusion of baryons substantially reduces any distinct signatures of dark matter physics in the observable properties of dwarf galaxies. Spatially-resolved rotation curves in the central regions ($<400$ pc) of small dwarfs could provide a way to distinguish between CDM, WDM, and SIDM, however: at the masses probed in this simulation suite, cored density profiles in dwarfs with small $\rh$ values can only originate from dark matter self-interactions. 
\end{abstract}


\section{Introduction}
\label{sec:intro}
Dwarf galaxies continue to be one of the few areas where the cosmological constant + cold dark matter (\lcdm) cosmological model has difficulties matching observations. Though the theory has been a resounding success in matching the large-scale structure of the Universe (e.g. \citealt{Springel:2005}), it is still beset by a handful of pernicious issues at the dwarf galaxy mass scale whose resolution may be dark matter that has properties different from the standard picture of being cold and collisionless (see \citealt{Bullock:2017} for a review). 

Each of the main challenges to \lcdm~on the scale of dwarf galaxies can be traced back to comparisons between results from dark matter-only (DMO) simulations and observations of dwarf galaxies. The core/cusp issue stems from the discrepancy between the cuspy central density profiles found universally in DMO simulations \citep{Navarro:1996b,Navarro:1997,Moore:1999,Klypin:2001,Navarro:2004,Diemand:2005} and the core-like DM profiles favored by kinematic observations of the rotation curves of disc galaxies or the velocity dispersions of certain dwarf spheroidals \citep{Flores:1994,deBlok:2001,Salucci:2001,Kuzio:2006,Spano:2008,Oh:2011,Oh:2015,Walker:2011}. Similarly, attempts to directly compare the abundance of dark matter (DM) subhalos around Milky Way-mass hosts in DMO simulations with luminous satellites of the actual Milky Way (MW) have found that the former outnumbers the latter by orders of magnitude, resulting in the `missing satellites problem' (MSP; \citealt{Klypin:1999,Moore:1999}; see also \citealt{Kauffmann:1993}). Finally, if one attempts to resolve the MSP by placing the brightest satellites in the most massive subhalos found around MW-mass hosts in DMO simulations, the result is a gross mismatch between the observed and predicted stellar kinematics of the dwarfs. While this too-big-to-fail (TBTF) problem was initially found in the satellites of the Milky Way \citep{Boylan-Kolchin:2011}, it is not unique to them: it has since been expanded to the satellites of Andromeda \citep{Tollerud:2014} and the field galaxies of the Local Group (LG; \citealt{Kirby:2014,Garrison-Kimmel:2014}) and beyond \citep{Papastergis:2015}.

The mismatches described above are often explained as the natural result of physics missing from the simulations, with the accompanying expectation that the introduction of self-consistent modeling of galaxy formation physics will reconcile \lcdm\ theory with observations  \citep{Navarro:1996a,Governato:2010,Weinberg:2013}. Specifically, the MSP (and related issue of missing dwarfs; \citealt{Zavala:2009,Klypin:2015}) can be understood as a natural consequence of cosmic reionization \citep{Bullock:2000,Benson:2002,Somerville:2002,Okamoto:2008} combined with environmental stripping \citep{Buck:2018,Fillingham:2018,RodriguezWimberly:2018} and disruption within the gravitational potentials of massive galaxies (\citealt{DOnghia:2010,Sawala:2017,Garrison-Kimmel:2017b}, though see \citealt{VandenBosch:2017}). Baryons have been also theorized to have a substantial effect on the inner structure of dwarfs. Through repeated outbursts of star formation (with the associated supernovae feedback), dwarfs are able to blow out their central baryons and induce rapid changes in the gravitational potential that ultimately remove dark matter from galaxies' centers \citep{Pontzen:2012,Madau:2014}. Recent hydrodynamical simulations of dwarfs have confirmed this behavior \citep{Chan:2015,Onorbe:2015,Read:2016a,Tollet:2016,Fitts:2017}, though this result is not universal  \citep{Chen:2016,Sawala:2016b}. Dissipative baryonic physics and selective disruption have also been shown to effectively lower the peak circular velocities of the most massive satellites in simulated LG pairings, leaving the simulations free of the TBTF problem \citep{Brooks:2014,Dutton:2016,Wetzel:2016,Garrison-Kimmel:2018}. 

While the effects of baryonic feedback in simulations are encouraging in their ability to match observations, lingering doubts remain about the necessity of baryon-induced core formation and the agreement between simulations and the wide range of properties in observed galaxies (see, e.g., \citealt{Oman:2015,Sawala:2016b,Sales:2016}). Moreover, cored profiles generated by feedback, as proposed in \citet{Dicintio:2014b,Dicintio:2014}, may be inconsistent with the correlations predicted in \lcdm~cosmology, namely the mass-concentration and $\mstar-\mhalo$ abundance matching relations (\citealt{Pace:2016}; however, see \citealt{Katz:2017}). Others have argued that cores may simply be observational artifacts \citep{Pineda:2017,Oman:2017}. Recent studies have also used the multiple stellar populations of Sculptor and Fornax to call into question the existence of cores in these systems (\citealt{Genina:2018}, though see \citealt{Hayashi:2018}'s work on Carina). Additionally, \citet{Papastergis:2016} found that abundance matching the observed rotation velocity function of HI gas of dwarfs from the ALFALFA survey still results in a TBTF problem. It remains an intriguing possibility that small-scale issues may not be solved by the simple inclusion of baryons in CDM simulation and may lie beyond the CDM paradigm \citep{Smith:2018}.

The difficulties inherent in making ab initio predictions in fully hydrodynamical simulations, coupled with the shrinking parameter space for WIMP-like dark matter \citep{Aprile:2018}, have led to significant explorations of dark matter models other than CDM (see, e.g., \citealt{buckley2018} for a recent review). One compelling alternative is a possible `warm' dark matter (WDM), in which free-streaming of dark matter erases primordial perturbations with masses below a model-dependent scale \citep{Bond:1982,Hogan:2000,Sommer-Larsen:2001,Bode:2001,Barkana:2001}. Initially introduced as a natural way to smooth out the inherently clumpy nature of \lcdm~ and thus address the Missing Satellites problem \citep{Colin:2000,Polisensky:2011,Lovell:2012,Anderhalden:2013,Bozek:2016,Horiuchi:2016}, WDM results in lower central densities within dark matter halos -- though the halos are still cuspy on scales relevant for observations of dwarf galaxies -- thereby addressing TBTF \citep{Lovell:2012,Horiuchi:2016,Lovell:2017}. 

In addition, WDM is also well-motivated as a potential source of the highly-debated detection of a 3.55 keV line in the X-ray flux observed in the center of the MW, M31, the Perseus cluster, and stacked observations of other clusters \citep{Boyarsky:2014,Bulbul:2014,Boyarsky:2015, Iakubovskyi:2015,Abazajian:2017}. The next step in testing WDM as a viable option has been including the effects of baryons in WDM. While some have already simulated the Local Group in WDM with a semi-analytical treatment of hydrodynamics \citep{Lovell:2016,Bose:2017}, others have simulated individual dwarfs in a cosmological context with full hydrodynamical treatment \citep{Governato:2015,Gonzalez-Samaniego:2016,Bozek:2018}. Both approaches have produced predictions to distinguish CDM dwarfs from WDM dwarfs -- e.g., reduced stellar masses  \citep{Gonzalez-Samaniego:2016} and purely young galaxies \citep{Bozek:2018} in WDM -- however, the limited work on this topic to date leaves the question far from settled.

Another compelling alternative is the possibility of a theory of dark matter that allows for self-interactions. Initially, self-interacting dark matter (SIDM) was invoked for its ability to produce constant-density cores in the center of DM halos through strong elastic self-scattering interactions \citep{Spergel:2000}. Simulations have since confirmed this ability with increasingly higher levels of resolution \citep{Dave:2001,Rocha:2013,Zavala:2013}, providing a new avenue to resolving the core/cusp issue. Other simulations, focused on low-mass galaxies with maximum circular velocity of $\vmax\simeq30$ km s$^{-1}$, have found that self-interaction cross section values of $\sigma/m=0.5$ to 10 cm$^2$ g$^{-1}$ at the scale of dwarf galaxies are able to solve the core-cusp and TBTF issues present in \lcdm\ \citep{Vogelsberger:2012,Peter:2013,Rocha:2013,Zavala:2013,Elbert:2015,Fry:2015}. The bulk of these results have not included the effects of baryons, however.

Those groups that have included hydrodynamics are now able to produce SIDM simulations broadly consistent with dwarf galaxies, though the predicted properties in the inner regions have been found to be mutually inconsistent. For example, both Vogelsberger et al.~(\citeyear{Vogelsberger:2014b}) and Fry et al.~(\citeyear{Fry:2015}) performed hydrodynamical simulations of dwarf galaxies in SIDM and found the majority (both isolated and not) to be nearly identical to the CDM versions at a radius of 500 pc.  However, Robles et al.~(\citeyear{Robles:2017}) recently simulated 4 dwarf galaxies (M$_\mathrm{vir}\approx 10^{10}\,\msun$, $\mstar\approx 4\times10^5-10^7\,\msun$)  at higher resolution and found a $\mstar$-dependent difference between the CDM and SIDM hydrodynamical versions in both density profile slopes and magnitudes at 500 pc. 

Baryons also introduce the prospect of gravothermal core collapse in SIDM simulations, leading to denser central regions than what is expected from SIDM alone \citep{Kochanek:2000, Balberg:2002, Colin:2002, Koda:2011, Vogelsberger:2012}. Initial simulations found this effect was possible only with cross-sections $\sigma/m\gtrsim10$ cm$^2$ g$^{-1}$ \citep{Elbert:2015}. However, simulations with semi-analytic treatment of the baryons, at a range of different halo mass scales, found that core collapse was possible with a cross-section of $\sigma/m=0.5$ cm$^2$ g$^{-1}$, so long as the stellar potential dominated the central parts of a galaxy \citep{Elbert:2018}. Any attempt to address the small-scale issues of \lcdm~ with SIDM will clearly require a comprehensive understanding of its interaction with baryons.

Given the viability of possible alternatives to \lcdm, it is important to study them side by side so as to gain a deeper understanding of the specific ways each theory affects galaxy properties and what those differences might mean for observations. Studies based DMO simulations have investigated whether different dark matter models (including WDM, `mixed' (warm+cold) dark matter, and SIDM) can remedy the mismatch between observational and simulated velocity functions \citep{Schneider:2017}. A fully consistent treatment of hydrodynamics and galaxy formation in such comparisons is still lacking. A central aim of this paper is to investigate the effects of baryonic physics in self-consistent simulations of non-CDM models. In order to focus on the scales relevant for TBTF and core/cusp, our suite is comprised of halos at the edge of where stellar feedback is effective at modifying halos' central dark matter distributions in CDM \citep{Governato:2012,Dicintio:2014b,Onorbe:2015,Chan:2015,Fitts:2017}, $\mhalo\sim10^{10}\,\msun$. These dwarfs are also selected to be isolated from any larger galaxies to provide a testing ground free from environmental processes.

The paper is arranged as follows. \S\ref{sec:simulations} provides a brief overview of the simulation suite, including the various DM theories considered. \S\ref{sec:results} outlines the main results of our study, including the mass assembly histories for the suite, a number of global properties of the simulated dwarfs (and how they compare to observations) as well as a dedicated look at central density profiles and rotation curves found in each version of DM. Our analysis in \S\ref{sec:discussion} places our simulations alongside observations in order to investigate any potential signatures of non-standard DM models and to understand the interplay of DM physics and baryonic physics in the simulations. We also compare to several of the latest attempts to simulate a realistic Local Field dwarf population. Finally, we summarize our results and conclusions in \S\ref{sec:conclusions}. We assume a background cosmology derived from the \textit{Wilkinson Microwave Anisotropy Probe} 7-year data \citep{Komatsu:2011}: $h = 0.71$, $\Omega_{\rm m} = 0.266$, $\Omega_{\rm b} = 0.0449$, $\Omega_\Lambda = 0.734$, $n_{\rm s}$ = 0.963, and $\sigma_8$ = 0.801.
\begin{figure}
\includegraphics[width=\columnwidth]{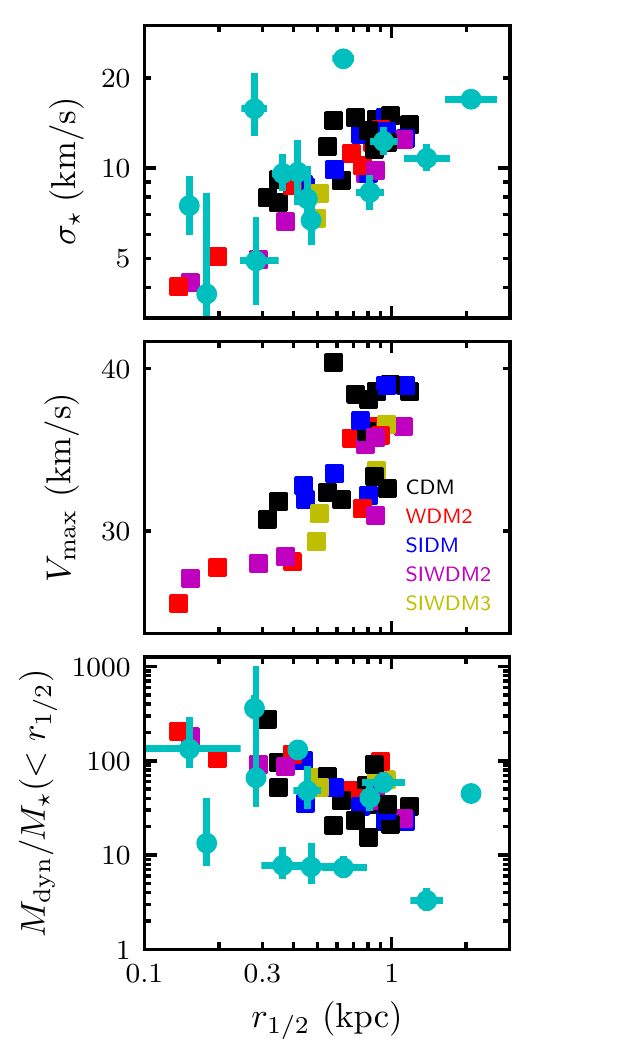}
    \vspace{-0.5cm}
  \caption{\textit{Top}: 1D stellar velocity dispersion (computed as $\sigma_{\star,\mathrm{3D}}/\sqrt{3}$) as a function of the 3D stellar half-mass radius $\rh$. \textit{Middle}:  Maximum of the circular velocity curve, $V_\mathrm{max}$ as a function of $\rh$; the vertical axis is plotted on a logarithmic scale. \textit{Bottom}:  Ratio of total (dynamical) mass to $\mstar$ within $\rh$ as a function of $\rh$. Simulated galaxies are plotted as squares and are colored according to their version of DM: CDM in black, WDM2 in red, SIDM in blue, SIWDM2 in magenta, and SIWDM3 in yellow. Data for low-mass dwarfs in the Local Field (as cyan circles, from \citealt{Garrison-Kimmel:2018}) are also plotted for comparison. In the top and bottom panels, the simulations follow the same trends as the observations and fall in the same part of parameter space. Each relation remains tight across all five different DM theories, with the halo-to-halo scatter in CDM exceeding the scatter originating about the average relation in non-CDM models.}
  \label{fig:trip}
\end{figure}
\begin{figure}
\includegraphics[width=\columnwidth]{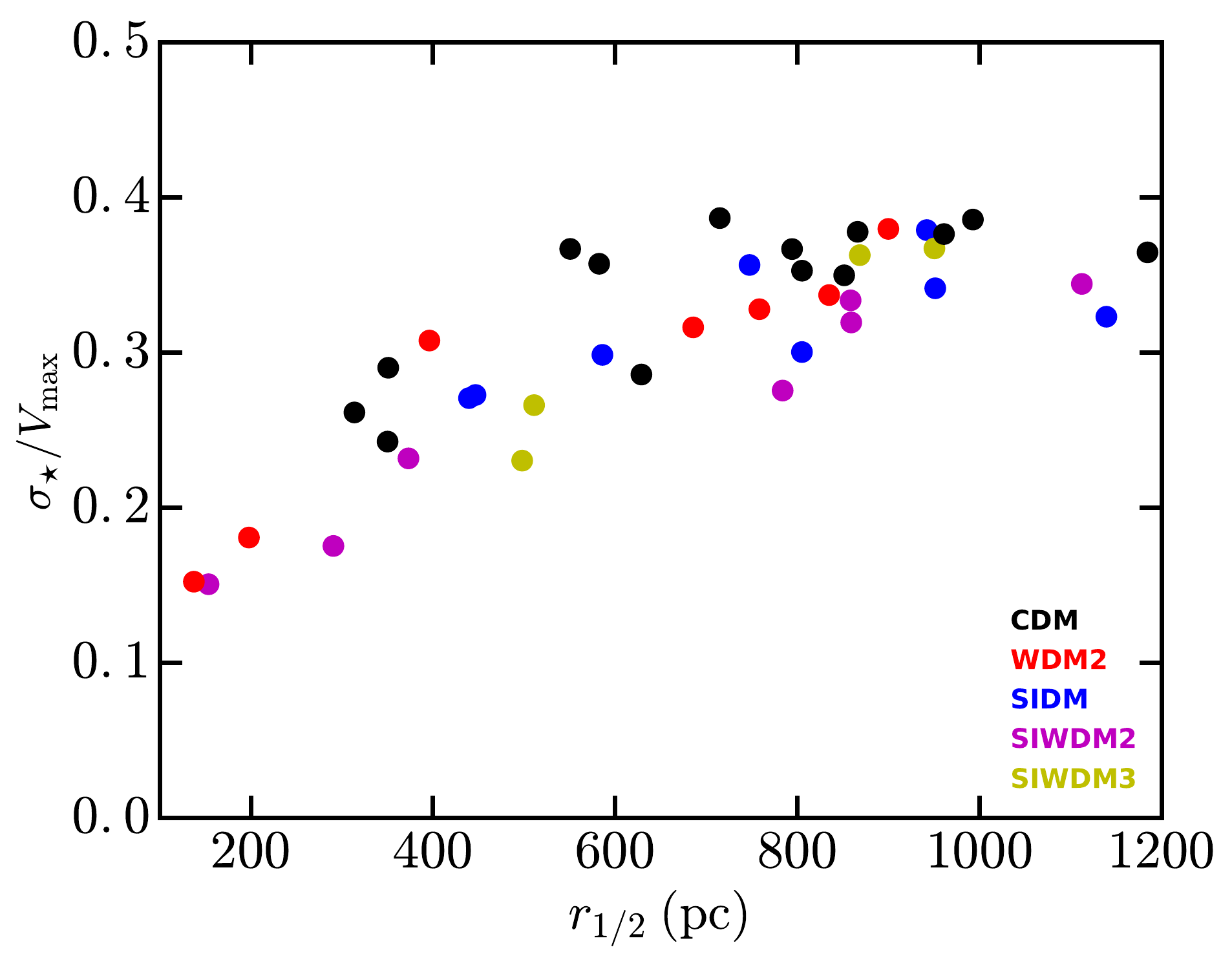}
    \vspace{-0.5cm}
  \caption{The ratio of each simulation's 1D stellar velocity dispersion to its maximum circular velocity as a function of $\rh$. Colors are identical to Fig. \ref{fig:trip} coloring scheme. For large $\rh$, there appears to be an upper limit to $\sigma_\mathrm{\star,1D}/\vmax$. The ratio becomes smaller in smaller systems, implying that observed galaxies with low $\sigma_\mathrm{\star,1D}$ and small $\rh$ could live in halos that are more massive than might naively be expected.}
  \label{fig:vel_ratio_vs_rhalf}
\end{figure}
\begin{figure}
\includegraphics[width=\columnwidth]{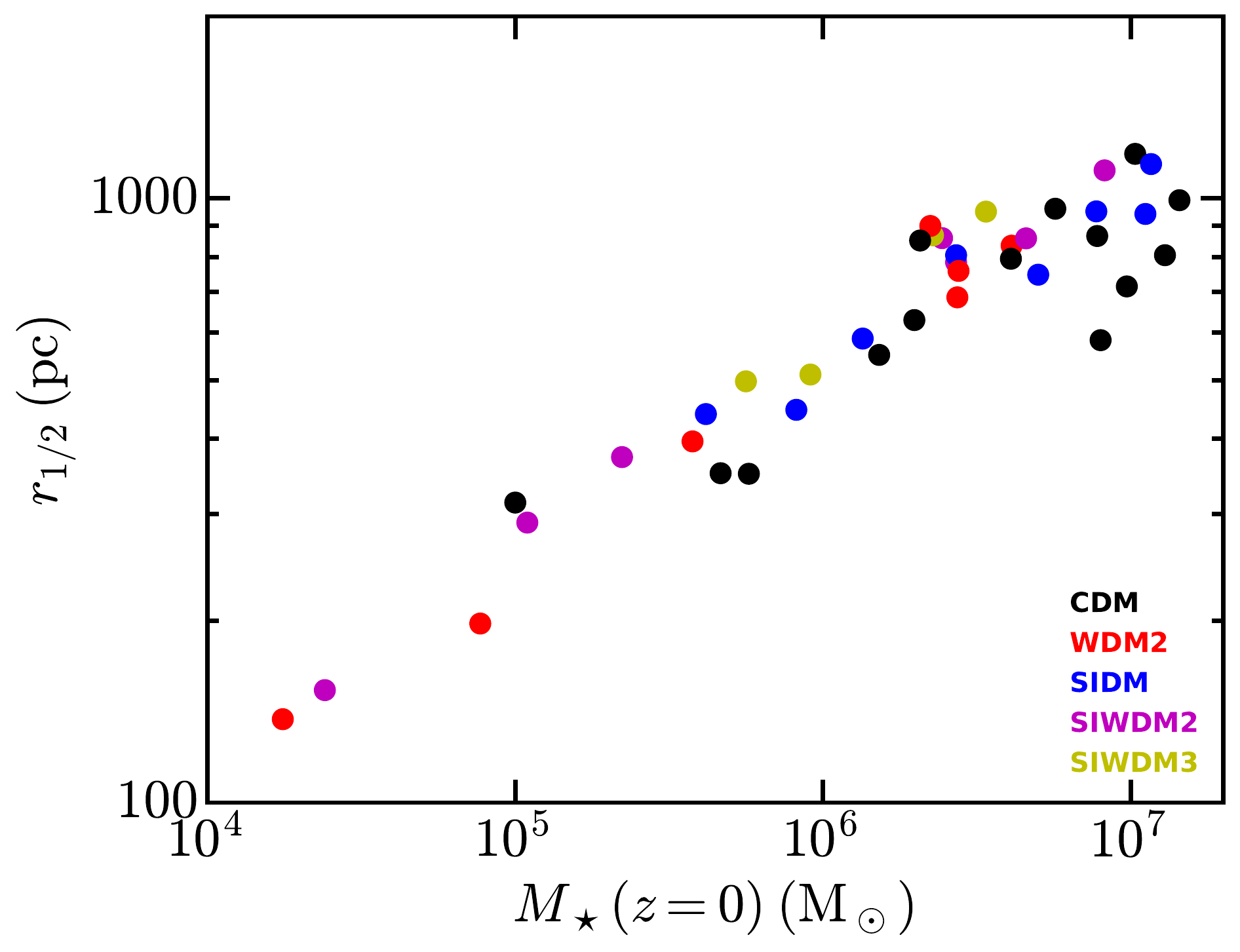}
    \vspace{-0.5cm}
  \caption{The correlation between $\mstar$ and $\rh$ for the entire simulation suite, across all 5 different types of DM. The connection between $\mstar$ and $\rh$ remains strong and consistent across all types of DM simulated here.}
  \label{fig:mstar_vs_rhalf}
\end{figure}
\begin{figure*}
\includegraphics[width=\textwidth]{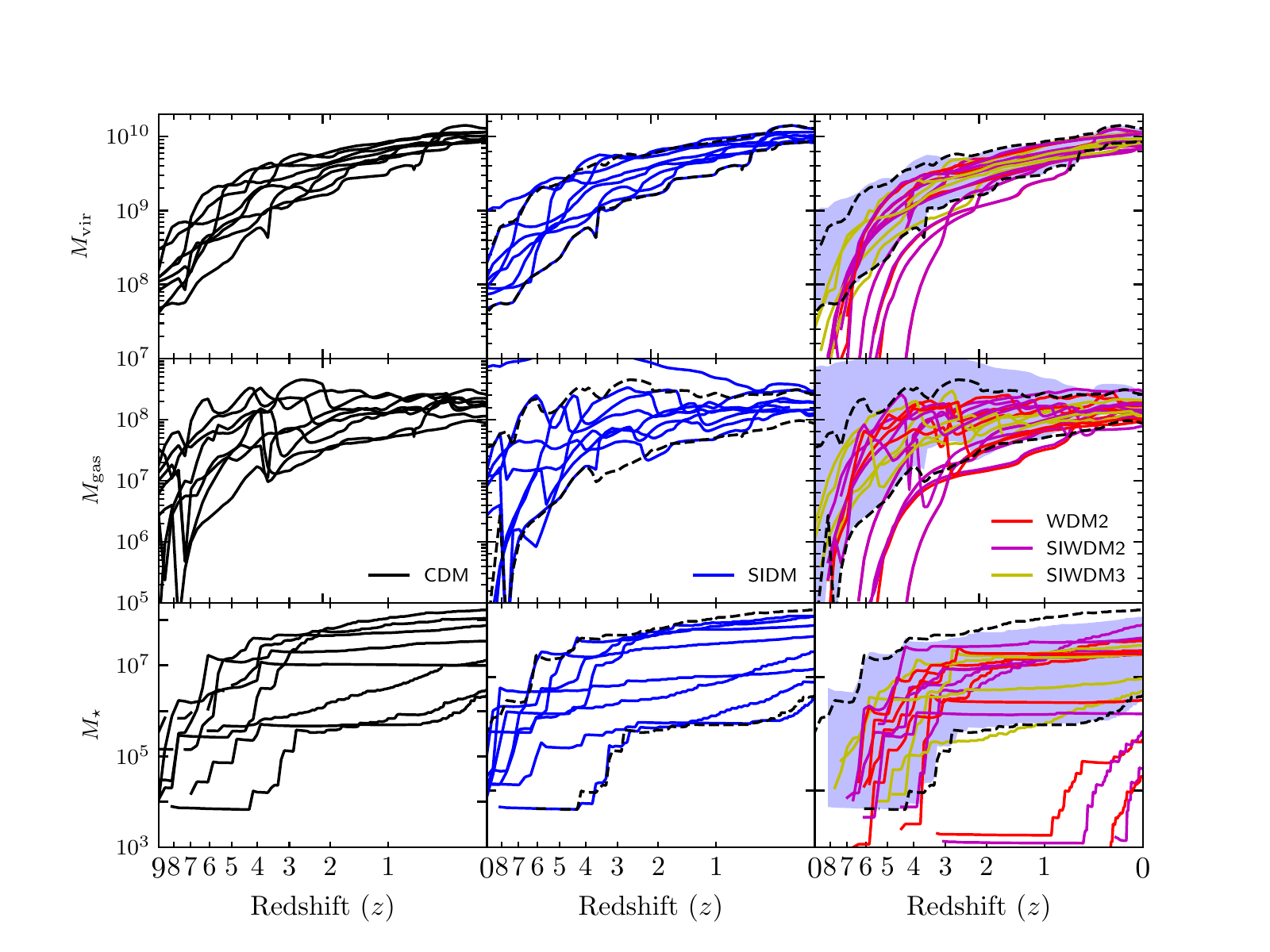}
    \vspace{-0.5cm}
  \caption{The different mass assembly histories in the hydrodynamical runs for the sub-sample of eight dwarfs simulated in various DM theories. The top row presents the virial mass of each dwarf. The middle row shows the total gas mass within the virial radius for all of the dwarfs. The bottom row displays the assembly histories of stellar mass within the inner galaxy ($<0.1\times R_\mathrm{vir}$) for all of the dwarfs. The original hydrodynamical simulations run in CDM are shown in the left column as black solid lines. To ease in the comparison of other DM versions of each dwarf, the range of these histories in the middle and right columns are marked by the dashed black lines. In the middle column we add the SIDM versions of each dwarf as blue solid lines. The rightmost column includes the WDM versions of dwarfs; the WDM2 is plotted as red, SIWDM2 as magenta and SIWDM3 as yellow.}
  \label{fig:mass_assembly}
\end{figure*}
\section{Simulations}
\label{sec:simulations}
Our simulation suite consists of 15 cosmological zoom-in simulations of \lcdm\ dwarf galaxy halos chosen to have virial\footnote{We define all virial quantities using the \citet{Bryan:1998} value of the overdensity $\Delta_{\rm vir}$. At $z=0$ in our chosen cosmology, $\Delta_{\mathrm{vir}}=96.45$ (relative to $\rho_{\mathrm{crit}}$), and $\mvir=10^{10}\,M_\odot$ corresponds to $R_\mathrm{vir}\approx56$ kpc.} masses of $10^{10}\,\msun \:(\pm 30\%)$ at $z=0$ (see \citealt{Fitts:2017} for details). The simulations here are part of the Feedback In Realistic Environments (FIRE, \citealt{Hopkins:2014})\footnote{\url{http://fire.northwestern.edu}}, specifically the ``FIRE-2'' version of the code; all details of the methods are described in \citep[][hereafter \hopkins, Section~2]{Hopkins:2018}. The simulations use the code GIZMO \citep{Hopkins:2015}\footnote{\url{http://www.tapir.caltech.edu/~phopkins/Site/GIZMO.html}}, with hydrodynamics solved using the mesh-free Lagrangian Godunov ``MFM'' method.  The simulations include cooling and heating from a meta-galactic background\footnote{The simulations used the "December 2011 update" of the FG09 model (available here: http://galaxies.northwestern.edu/uvb/), calibrated to produce a reionization optical depth consistent with WMAP-7 (corresponding to $z_{\rm reion}\sim 10$).} and local stellar sources from $T\sim10-10^{10}\,$K; star formation in locally self-gravitating, dense, self-shielding molecular, Jeans-unstable gas; and stellar feedback from OB \&\ AGB mass-loss, SNe Ia \&\ II, and multi-wavelength photo-heating and radiation pressure; with inputs taken directly from stellar evolution models. The FIRE physics, source code, and all numerical parameters are {\em exactly} identical to those in \hopkins. The fiducial simulations with galaxy formation physics included have baryonic (dark matter) particle masses of $500\,\msun$ ($2500\,\msun$), with a minimum physical baryonic (dark matter) force resolution of $h_{b}=2\,$pc ($\epsilon_{\mathrm{DM}}=35\,$pc); force softening for gas uses the fully-conservative adaptive algorithm from \citet{Price:2007}, meaning that the gravitational force assumes the identical mass distribution as the hydrodynamic equations (resulting in identical hydrodynamic and gravitational resolution). In post-processing, we identify halos and construct merger trees with the Amiga Halo Finder (\ahf; \citealt{Knollmann:2009}).

Eight of the halos (m10b, c, d, e, f, h, k, and m) were simulated in a WDM cosmology and were presented first in \citet{Bozek:2018}. The underlying dark matter particle model is a resonantly-produced sterile neutrino \citep{Shi-Fuller:1999} with a mass of $m_s = 7.1$ keV, a mixing angle of $\sin^2(2\theta) = 2.9 \times 10^{-11}$ and has a half-mode mass comparable with a thermal WDM model with $m_\mathrm{THM}=2$ keV. This model was selected to (1) provide free-streaming effects that are at the edge (i.e., the warmest) of what is allowed based on satellite galaxy counts and large-scale structure constraints of the Lyman-$\alpha$ forest and (2) account for the origin of possible detections of an X-ray line at 3.55 keV in galaxy and galaxy cluster observations. This allows us to test the strongest free-streaming effects possible given the current observational constraints.

The same eight halos were simulated again using a CDM power spectrum but with a self-interaction cross section of $\sigma/m=1\:\mathrm{cm}^2\,{\rm g}^{-1}$ using the SIDM implementation of \citet{Rocha:2013}. Four of these halos -- m10b, d, f and k -- were previously presented in \citet{Robles:2017}. We select this particular cross section as N-body simulations have converged on a $\sigma/m\approx0.5-1$ cm$^2\,{\rm g}^{-1}$ to solve the core-cusp and TBTF issues on small scales while remaining within the constraints from larger scales. Recent studies at the massive cluster scale have favored an even smaller cross section around 0.1 cm$^2\,{\rm g}^{-1}$ \citep{Kaplinghat:2016,Elbert:2018}, possibly pointing to a velocity-dependent cross section that would allow for smaller cross sections at more massive scales (see \citealt{Tulin:2018} for a review of current SIDM simulations and constraints). The choice of a constant cross section does not exclude this possibility at higher mass scales and is effectively equivalent within our narrow mass scale.  

Given that WDM and SIDM are both allowed by current data and have somewhat different effects on dwarf galaxy formation, we also look at a combination of the two for the same eight dwarfs. Our motivation is to understand the coupled effects of WDM (free-streaming and delayed structure formation) and SIDM (density reduction in halo centers). Finally, for a sub sample of the eight halos (m10d, e, f and k), we simulated a combination of SIDM with a slightly colder WDM (with half-mode mass equivalent to a thermal WDM model with $m_\mathrm{THM}=3$ keV). This slightly colder WDM is not only a sweet spot for the 3.55 keV decay signal \citep{Shi-Fuller:1999,Abazajian:2017} but may also be in better agreement with observations than our default $m_\mathrm{THM}=2$ keV model.  

\section{Results}
\label{sec:results}
Fig. \ref{fig:trip} shows various relationships for the suite of dwarfs: the one-dimensional stellar velocity dispersion (calculated as $\sigma_{\mathrm{1D},\star}=\sigma_{\mathrm{3D},\star}/\sqrt{3}$ based on all of the stars within each galaxy; top), the maximum circular velocity (middle), and the ratio of dynamical mass to stellar mass within stellar half-mass radius (bottom, with dynamical mass being the sum of baryonic and dark matter mass) as a function of the stellar half-mass radius, $\rh$. The simulations are represented as squares, colored according to dark matter variant. For comparison, we show data for low-mass dwarfs in the Local Field (defined here as within 1 Mpc of the MW or M31, but more than 300 kpc from both), compiled in \citet{Garrison-Kimmel:2018}, as cyan circles. These dwarfs span $10^5-10^8\,\msun$ in stellar mass and hence serve as a reasonable comparison to the simulations. Overall, the simulations produce a fairly tight grouping across all DM theories and are generally consistent with the population of dwarfs in the Local Field. All three relations are resilient to changing both the free-streaming length of the DM particle as well as including the possibility of DM self-interactions. 

The top two panels viewed together are particularly interesting, as the mapping between the observable ($\sigma_\star$) and the ``theory'' quantity ($\vmax$) is generally unknown. For a equilibrium dispersion-supported system, we expect that $\vmax \geq \sqrt{3}\,\sigma_\star$ or $\sigma_{\star} \leq 0.577\,\vmax$ (e.g., \citealt{Wolf:2010}), with the maximal value attained if the galaxy size is identical to the radius where the peak circular velocity is attained ($r_{\rm max}$). Indeed, if we plot the ratio of $\sigma_\star/\vmax$ in Fig. \ref{fig:vel_ratio_vs_rhalf} (with colors identical to Fig. \ref{fig:trip} coloring scheme) as a function of $\rh$ we see that for large $\rh$, there is an upper limit to $\sigma_\star/\vmax$ of $\sim 0.4$, and this upper limit is well below the theoretical maximum of $\sim 0.58$. Even more interesting is that when looking at smaller systems, this ratio gets smaller; this implies that galaxies with small $\sigma_\star$ and small $\rh$ could live in relatively more massive halos than might be naively inferred from their kinematics, which has implications for the MSP and the TBTF problem. These results appear to hold across all DM theories tested here. If this relationship holds over a wider range of halo masses, 
it would prove very useful in matching observed galaxies to simulated halos. 

Despite the previous relations holding across multiple theories of dark matter, we find systematic effects in halo and galaxy properties when transitioning from one DM theory to another. For instance, increasingly warm theories of DM result in galaxies that are systemically smaller in size than their CDM counterparts, with an accompanying reduction in stellar mass. Moving from CDM to WDM2 shrinks $\rh$ in all 8 dwarfs by 40$\%$ on average, with correspondingly smaller stellar masses, keeping the galaxies on the same $\rh-\mstar$ relation. Meanwhile, including self-interactions into CDM does not affect $\rh$ or $\mstar$ significantly, with an average increase in $\rh$ of $10\%$ (and the most extreme increase being an increase only $28\%$). Introducing self-interactions into WDM2 on average also increases the dwarfs' sizes on average by $22\%$, though this still places them at roughly $70\%$ the size of their CDM counterparts. Each dwarf simulated in SIWDM3, the `lukewarm' DM between CDM and WDM2, reaches $90\%$ the size of its CDM versions, intermediate to the results of CDM and WDM2. Moving to a warmer theory of DM generally correlates with a decrease in $\rh$ and $\mstar$, while including self-interactions into CDM or WDM2 can make both quantities slightly larger. In no cases do we find systems that match the enigmatic systems of Crater 2 \citep{torrealba2016} or Antlia 2 \citep{torrealba2018}, which have very large sizes and low velocity dispersions relative to their stellar masses when compared to other systems, in $\rh-\vcirc-\mstar$ space. Environment might have played an important role in the evolution of these galaxies; alternately, dark matter with properties different from any of the models considered here might be responsible for their unique nature.

Given the systematic changes in the sizes the simulated dwarfs experience when transitioning to DM models with larger free-streaming lengths, one might wonder whether the relationship between size and stellar mass may change along with it.  Previous \fire-2 studies of dwarf galaxies simulated using either SIDM \citep{Robles:2017} or WDM \citep{Bozek:2018} have found identical $\mstar$-$\rh$ relations compared to CDM. In Fig.~\ref{fig:mstar_vs_rhalf}, we see that both using a warmer theory of DM and/or including self-interactions leaves the relation intact for the extended suite of halos. The tightness of this correlation appears to be fundamentally tied to hydrodynamics and the global gravitational potential, an important point to which we will return in \S\ref{sec:discussion}. 

\begin{figure*}
\includegraphics[width=\textwidth]{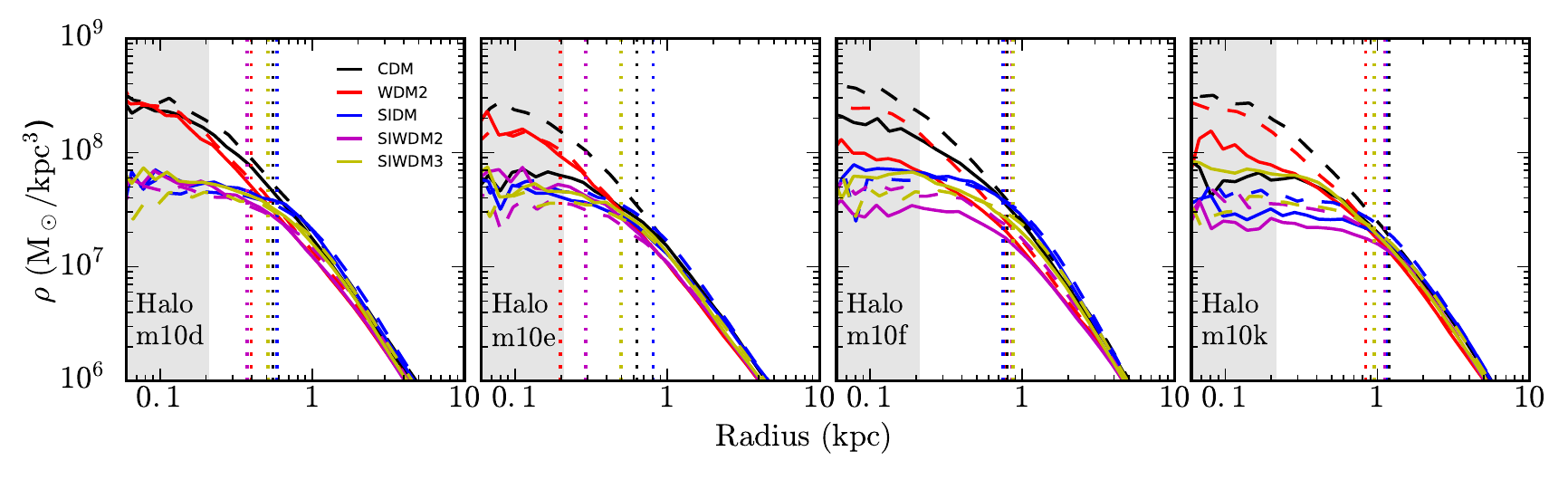}
    \vspace{-0.5cm}
  \caption{Radial density profiles for the 4 dwarfs in the suite simulated in all five different types of DM (following the color convention of Fig. \ref{fig:trip}) with stellar mass increasing from left to right. The dashed lines represent the DMO version of each simulation while the solid lines represent the hydrodynamical versions. The grey shaded region shows where numerical relaxation may affect the CDM density profiles according to the Power et al.~(\citeyear{Power:2003}) criterion. Stellar half-mass radii for each DM version are shown as vertical dotted lines. Any variation between the density profiles, whether from baryonic feedback or self-interactions, is bound within $\rh$. In Halo m10d we see a clear distinction in the central region between those versions that do and don't have self-interactions. As we move to the right, this distinction becomes increasingly muddied as baryonic feedback has a larger impact on the runs without self-interactions.}
  \label{fig:radden_1x4}
\end{figure*}
\subsection{Assembly history in different DM theories}
Figure \ref{fig:mass_assembly} displays the different mass assembly histories in the hydrodynamical runs for the sub-sample of 8 dwarfs simulated in various DM theories. The top row shows the evolution of the virial mass of each dwarf, the middle row shows the total gas mass within the virial radius, and the bottom panel shows stellar mass of the main progenitor within the inner galaxy ($<0.1\times R_\mathrm{vir}$). The original hydrodynamical simulations run in CDM are shown in the left column as black solid lines. In order to simplify comparisons across DM variants, the range of these histories in the middle and right columns are marked by the dashed black lines. In the middle column we add the SIDM versions of the dwarfs as blue solid lines. Again, to ease in comparison, the range of SIDM histories is plotted as a shaded blue region in the rightmost column. 

As expected, moving from CDM to SIDM has little effect on the DM mass assembly and ultimately results in very similar assembly histories, as is evident in the upper middle panel. With the exception of one dwarf (m10m, which has 50$\%$ less $\mstar$ at $z=0$ than its CDM counterpart), SIDM also has little effect on M$_\mathrm{gas}$($z=0$) and shows a nearly identical range as CDM for $\mstar(z)$. Despite the majority of star formation happening in situ for dwarfs of this mass range \citep{Fitts:2018}, the lower central densities appear to have little effect on the overall star formation history and ultimately $\mstar(z=0)$ of isolated dwarfs in the $10^{10}\,\msun$ mass range. Similar to what was found in Fry et al.~(\citeyear{Fry:2015}), the baryonic assembly histories in SIDM generally do not vary notably from their CDM counterparts, indicating they are tightly linked to the underlying DM assembly history.
\begin{figure*}
\includegraphics[width=\textwidth]{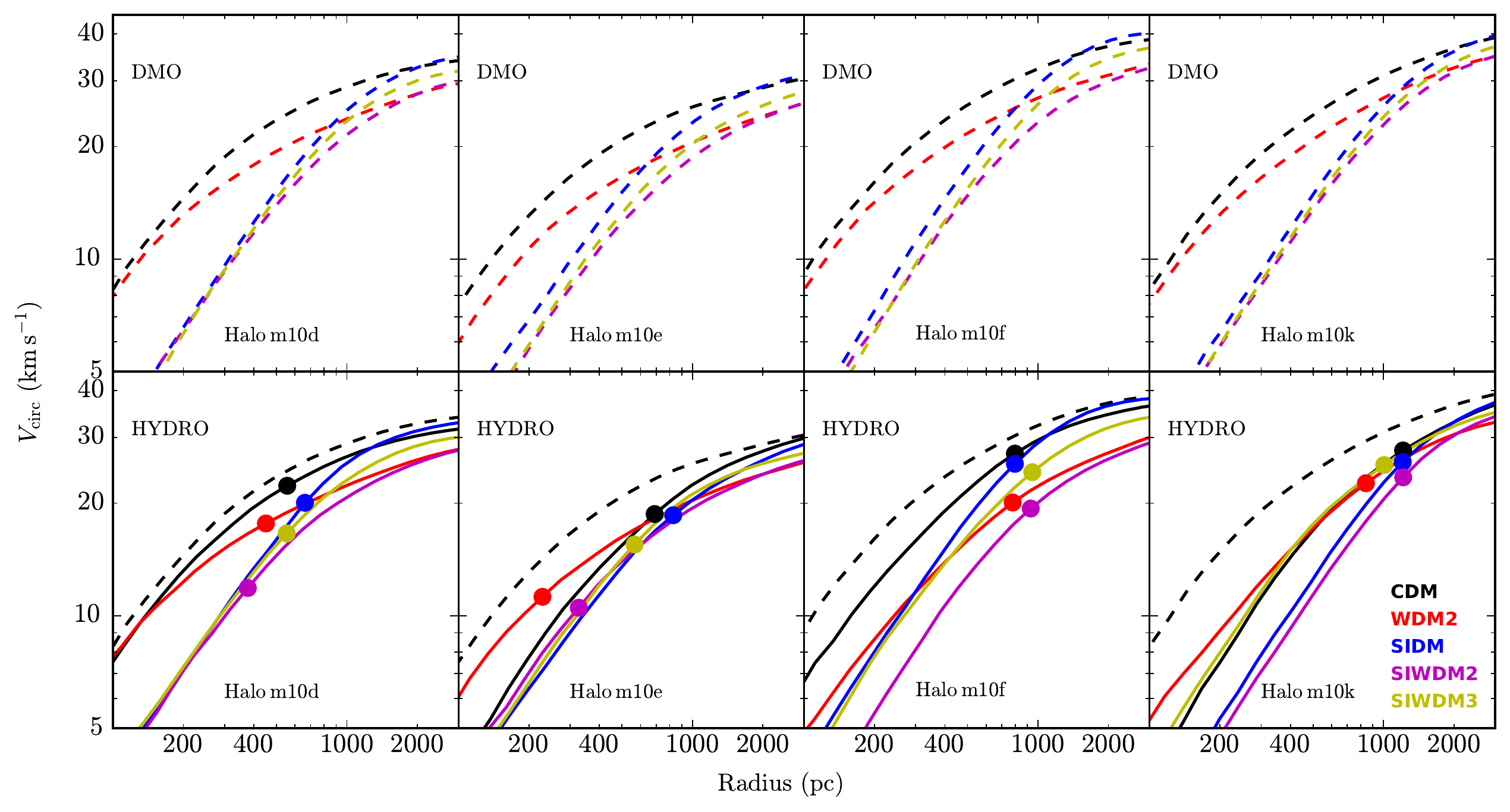}
    \vspace{-0.5cm}
  \caption{Rotation curves for the same 4 dwarfs present in Fig. \ref{fig:radden_1x4}. DMO simulations are represented as dashed lines and plotted in the top row while hydrodynamic simulations are represented as solid lines and plotted in the bottom row. The curve for the CDM DMO run is also included in the bottom row for reference. The circular velocity at $\rh$, $\vh\equiv V_\mathrm{circ}(\rh)$, for each curve is marked by a point with matching color. In every halo we see that including self-interactions provide an effective way to lower $\vh$; some halos even accomplish this with little change to $\rh$.}
  \label{fig:vcirc_1x4}
\end{figure*}
\begin{figure*}
\includegraphics[width=\textwidth]{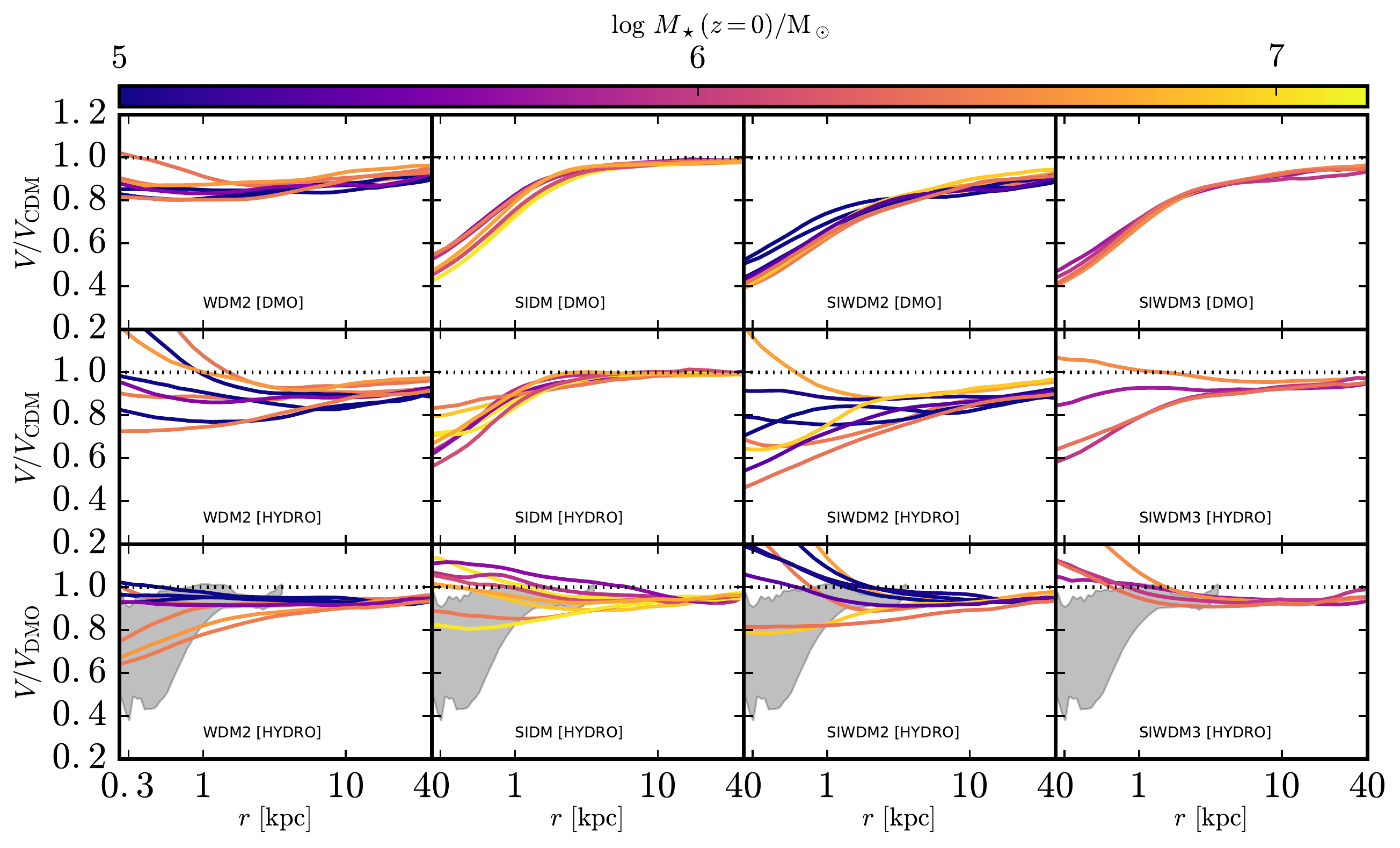}
\caption{\textit{Top Row}: Ratios of DMO rotation curves between each alternative DM theory and their CDM counterpart. Moving from left to right, each panel corresponds to WDM2, SIDM, SIWDM2 and SIWDM3 (each column is labeled in the bottom row). All lines are colored according to the stellar mass of their hydrodynamical counterpart. For WDM2, we see a broad 10-20$\%$ lowering of the rotation curve at all radii. Meanwhile by including self-interactions in the simulations we see that the inner kpc of the rotation curve is lowered. The two SIWDM columns show a mix of both properties, with the inner kpc lowered $40-60\%$ while the rest of the curve is only reduced a modest $\sim10\%$. \textit{Center Row}: Same as the top panel but now for the hydrodynamical simulations that include baryons. Now each effect noted in the previous row is not uniform across the dwarfs and nearly every dwarf sees less of a reduction when compared to it's CDM counterpart. \textit{Bottom Row}: Now to isolate the impact of hydrodynamics on each theory of DM we show the ratio of the hydrodynamical runs to their respective DMO runs. To understand the comparative effectiveness of hydrodynamical feedback in different DM theories, we overplot the lines on top of the range of ratios from the CDM runs (shown as the grey shaded region). If we focus solely on the WDM simulations (lower left panel), we see that the inclusion of baryons lowers the inner rotation curve compared to their DMO versions for those 3 dwarfs with the highest $\mstar(z=0)$ while changing little in those that form below $\mstar\sim10^6\,\msun$. If we focus on the simulations with self-interactions, the majority of dwarfs instead see a condensing of the inner kpc of the dwarf due to baryonic contraction of the central baryons. What allows a minority of the self-interacting dwarfs from preventing this contraction is not clear and does not appear to be linearly correlated with $\mstar(z=0)$.}
\label{fig:vcirc_ratio_3x4} 
\end{figure*}
\begin{figure}
\includegraphics[width=\columnwidth]{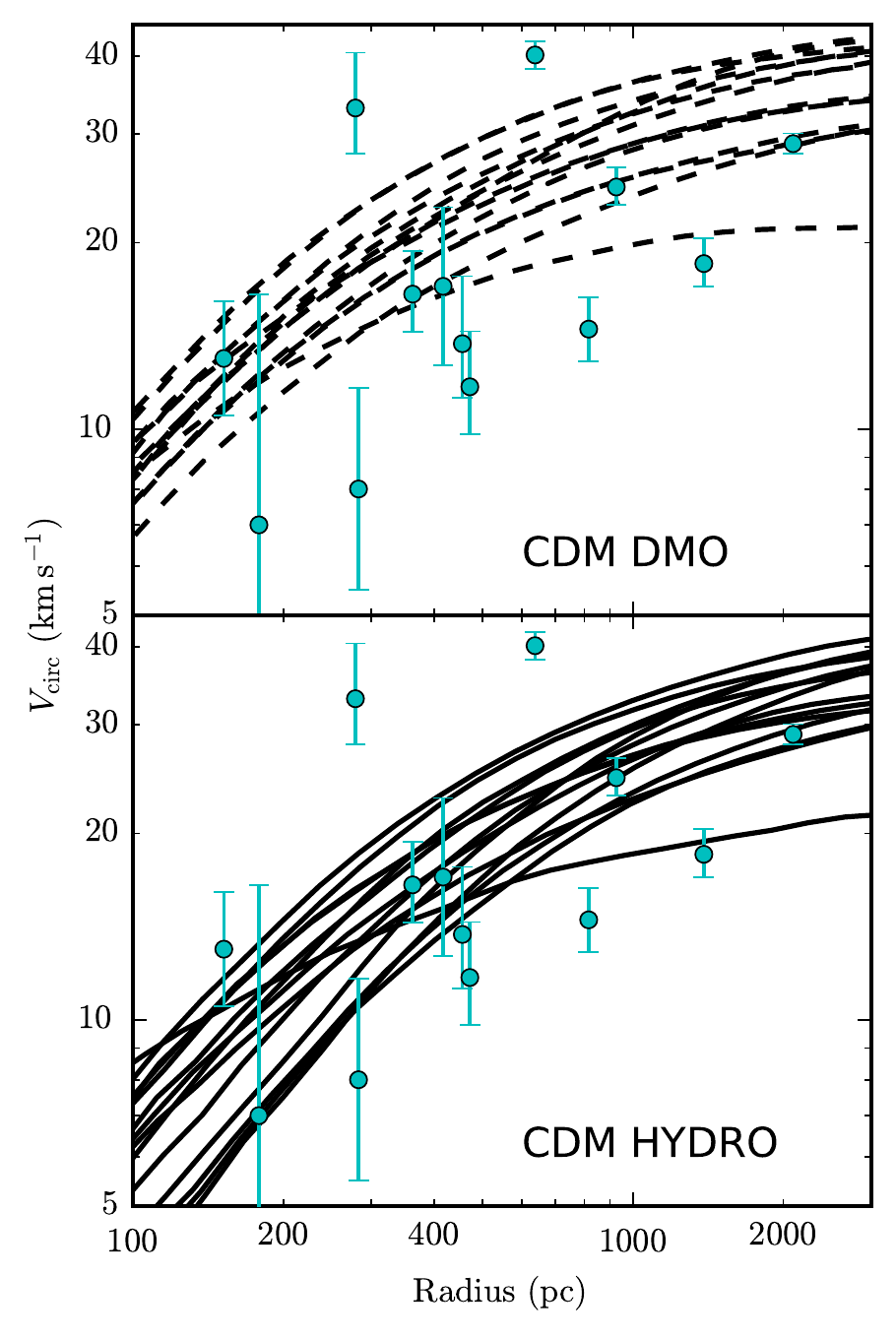}
    \vspace{-0.5cm}
  \caption{\textit{Top}: Rotation curves for all 14 DMO CDM dwarfs. $\vh$, the observed circular velocity at $\rh$, for Local Group field dwarfs are marked as cyan points. \textit{Bottom}: The same as the top panel but now for the hydro version of each CDM dwarf. If we focus only on the simulations' ability to produce rotation curves consistent with the observed points, the curves for the DMO simulations match only $\sim30\%$ of the points. The simulations specifically have trouble matching those with lower $\vh$ as well as the densest observed points. The addition of hydrodynamics further improves the CDM dwarfs' ability to match the observed low $\vh$ points, though this picture is not so clear if we additionally require the simulations to simultaneously match the observed dwarfs' size along with its $\vh$ (see Fig. \ref{fig:vcirc_2x4}). Also, the addition of hydrodynamics does little to address the two densest observed points (which prove difficult to explain with any of the simulations, see \S~\ref{sec:discussion}).}
\label{fig:vcirc_cdm}
\end{figure}
\begin{figure*}
\includegraphics[width=\textwidth]{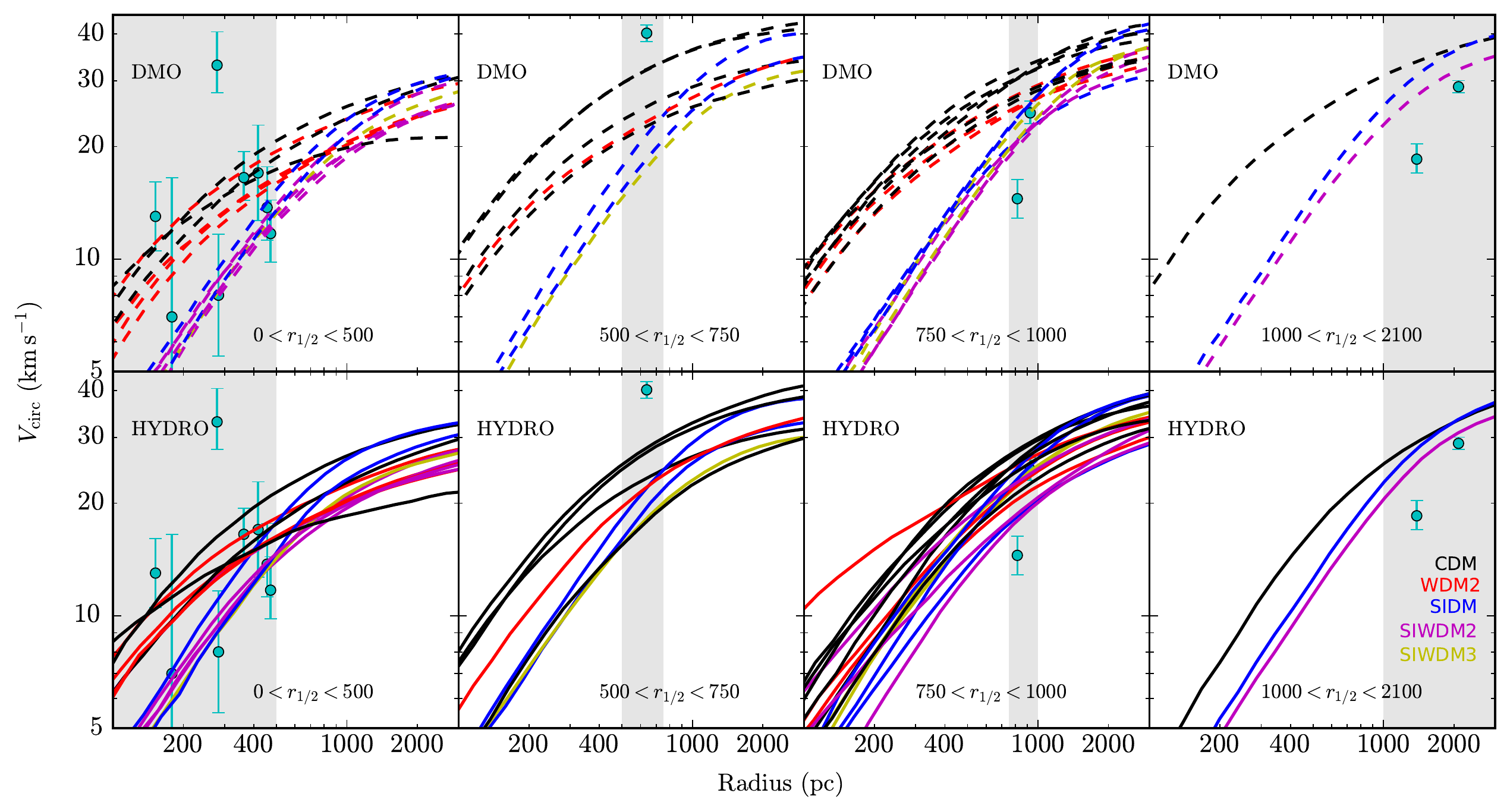}
    \vspace{-0.5cm}
  \caption{Rotation curves for the simulated dwarfs binned by size. The range of sizes for each bin is indicated by the shaded region in each subplot. The dashed lines in the top row are DMO simulations while the solid lines in the bottom row are the dwarfs run with hydrodynamics. The size of each dwarf in the DMO simulations is taken from the corresponding dwarf run with hydrodynamics. The colors follow the convention in Fig. \ref{fig:radden_1x4}. $V_{1/2}$ (the velocity at $r_{1/2}$) for observed Local Group field dwarfs are plotted as cyan points. While initially the CDM DMO simulations were consistent with $75\%$ of observed points (excluding the very dense Tucana), if we add the additional constrain of matching the observed dwarf's $\rh$, the CDM DMO simulations are only consistent with $50\%$ of observed points. Similarly, the CDM dwarfs with hydrodynamics originally fit 92$\%$  of field dwarfs but are now only consistent with $58\%$. Including some form of self-interactions, specifically the SIDM (blue) and SIWDM2 (magenta) presents a better picture with only 3 observed dwarfs lacking any simulated equivalent.  We note that this result is not merely a quirk of $\rh$ binning, it remains nearly identical if we instead bin the simulations by their corresponding $\mstar$ as well.}
  \label{fig:vcirc_2x4}
\end{figure*}
\begin{figure*}
\includegraphics[width=0.95\textwidth]{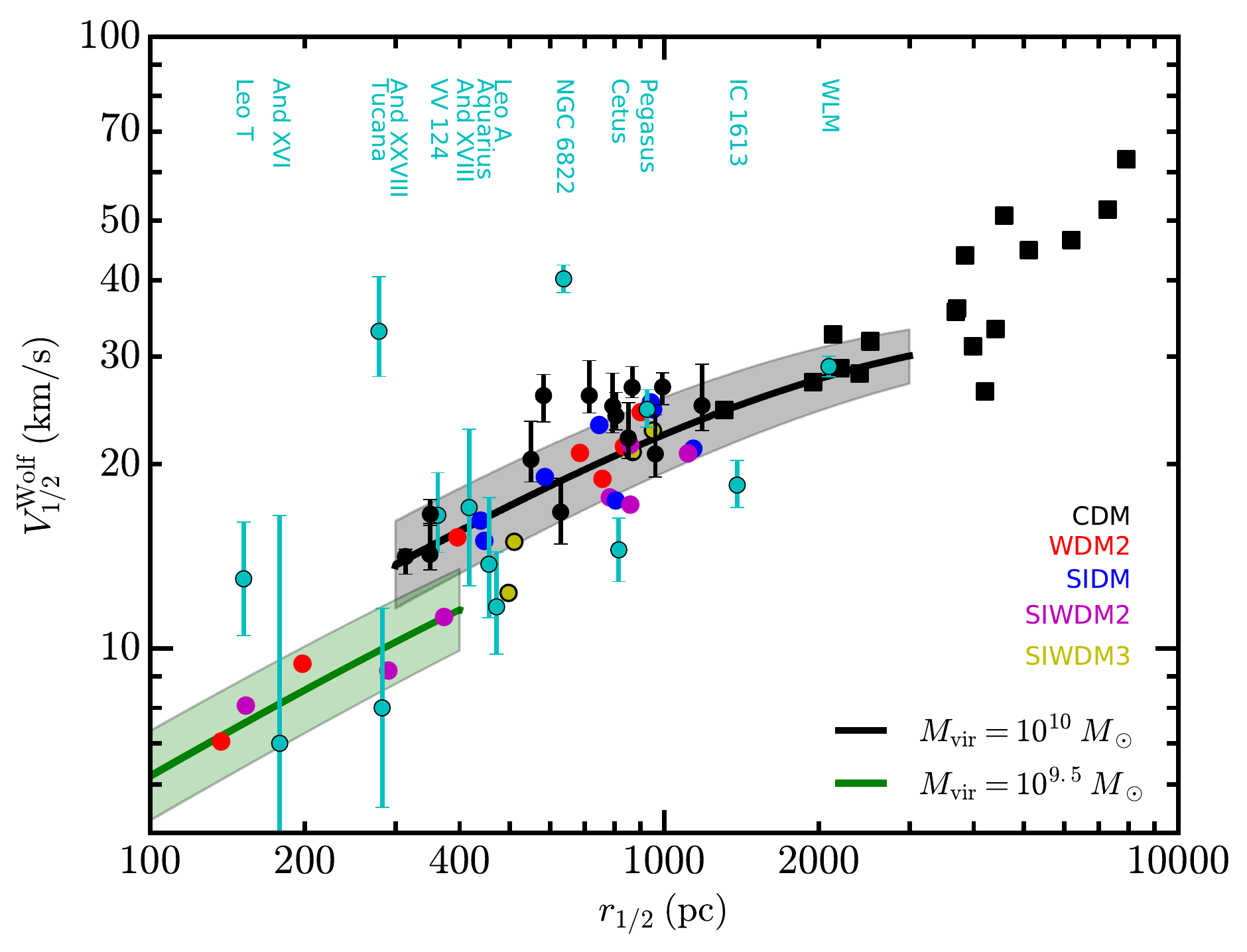}
\caption{$\vh-\rh$ relation for the suite of halos in various DM theories. $\mh$ for each halo is calculated using Eq.~\ref{eq:wolf} over 1000 random line-of-sight projections. Error bars on the CDM points indicate the 16th and 84th percentiles in the distributions. Cyan points denote observed $\vh$ values of Local Field galaxies. Black squares are data from the {\tt FIRE-2} simulations of isolated dwarfs with $10^{10}< \mh<10^{11}\,\msun$ from Graus et al.~(in prep.). We also include lines corresponding to the $\vh-\rh$ relation for a coreNFW profile \citep{Read:2016a}, with the the shaded region indicating the scatter in the $c-\mh$ relation (plotted only over $\mh$ values that are consistent with our $\mstar-\rh$ and $\mstar-M_\mathrm{halo}$ relations). All of the DM models simulated here follow the same median relation, to first order. While the CDM simulations do not produce any matches for the low $\rh$ and low $\vh$ galaxies (Leo T, And XVI and And XXVIII), the properties of these galaxies are in line with expectations for $\mvir\sim10^{9.5}\,\msun$ halos in CDM (according to the analytic coreNFW fits in green). Even accounting for differing halo masses, concentrations, the effect of mock observations, and different underlying DM models, the simulated suite of dwarfs still has mild difficulty matching the low $\vh$ Local Field dwarfs with $\rh>$500 pc (Cetus and IC 1613) and greater difficulty matching the Local Field dwarfs with high $\vh$ values and $\rh<600$ pc (NGC 6822 and Tucana).}
\label{fig:vhalf_vs_rhalf}
\end{figure*}

The rightmost column includes the WDM versions of dwarfs; the WDM2 is plotted as red, SIWDM2 as magenta and SIWDM3 as yellow. For increasingly large free-streaming lengths (from CDM to SIWDM3 to WDM2/SIWDM2), the fast accretion phase of mass assembly occurs increasingly later. Though the WDM dwarfs are quick to catch up to their CDM counterparts, the simulated dwarfs' $\mvir$($z=0$) are anywhere between $4-22\%$ smaller in WDM2/SIWDM2. The dwarfs' M$_\mathrm{gas}$($z=0$) are generally smaller (by as much as 50$\%$) in WDM2 relative to CDM. We also see significant reduction in the stellar mass compared to their CDM versions: 3 out of the 8 WDM2 galaxies have $<10\%$ of the stellar mass their CDM counterparts at $z=0$. Though $\mstar(z=0)$ is strongly correlated with the $\vmax$ of the WDM halos (similar to what is found in CDM, \citealt{Fitts:2017}), it is difficult to know a priori from the CDM halo or galaxy properties which halo will suffer greater reduction in $\mstar$ when resimulated in a DM model with a larger free-streaming length. None of the masses at $z=0$ ($\mvir$, M$_\mathrm{gas}$ nor $\mstar$) correlate with the reductions, nor does $\vmax$ prove to be a useful indicator of reduction.

Including self-interactions in WDM does not appear to have an additional systematic effect on the halos' assembly histories.  Star formation is delayed in the two halos that assemble late and form few stars, however. The WDM2 version of Halo m10e (the second-lowest red line in the bottom right corner panel) has an additional 3 Gyr delay in its star formation when self-interactions are introduced (the second-lowest magenta line in the same panel). Halo m10c only forms $\sim2.5\times 10^{4}\,\msun$ of stars in WDM2, all very late in its lifetime; in SIWDM2, it does not form any stars. It contains over $2\times10^6\,\msun$ of HI gas at $z=0$, however, indicating that if the simulation were continued into the future, it would likely form stars. It is not clear whether self-interactions alone are responsible for these delays in star formation, though, as the same behavior is not present when introducing self-interactions into CDM simulations. m10b, the halo that forms the fewest stars in CDM, shows the opposite behavior when including self-interactions: it fails to form any stars in WDM2 (though has HI gas within the virial radius at $z=0$), while in the SIWDM2 version it has a brief burst of very late star formation. It is also important to note that galaxy formation at this mass scale is inherently stochastic, with run-to-run variation influencing the exact mass assemblies. Overall, it appears that introducing self-interactions into either WDM or CDM simulations has little (additional) effect on the assembly histories. 

\subsection{Density profiles}
The dark matter density profiles of the 4 dwarfs simulated in all five different types of DM are plotted in Fig.~\ref{fig:radden_1x4}. The dashed lines represent the DMO version of each simulation while the solid lines represent the hydrodynamical versions. The grey shaded region indicates where numerical relaxation may affect the CDM density profiles according to the Power et al.~(\citeyear{Power:2003}) criterion. The stellar mass of the halos for a given dark matter model generally increases left to right. The only simulations that do not follow this behavior are the three versions of Halo m10e that are run with a version of WDM (WDM2, SIWDM2 or SIWDM3). Each of these dwarfs forms fewer stars than the corresponding version for Halo m10d, which may be because each WDM or SIWDM version Halo m10e forms its stars later (considerably so in the case of WDM2 and SIWDM2).

All of the simulations with self-interactions yield cores in their density profiles. However as one looks at dwarfs of increasingly higher stellar mass, it becomes increasingly difficult to distinguish the profiles of separate types of dark matter. In Halo m10d (the leftmost panel in Fig.~\ref{fig:radden_1x4}), we see a clear distinction in the central region between those versions that do and don't have self-interactions. As we move to the right, however, this distinction becomes increasingly muddied as baryonic feedback has a larger impact on the runs without self-interactions.

The effects of galaxy formation do not lead to an equal reduction in the inner dark matter density for all of the dark matter variants considered here. In \citet{Fitts:2017}, all CDM dwarfs with $\mstar>2\times10^6\,\msun$ saw significant reduction of the central density (see their Fig. 7) compared to their DMO counterparts.  \citet{Bozek:2018} found that while the WDM simulations with the inclusion of hydrodynamics generally resulted in additional reduction of the inner dark matter density, feedback-related density reduction was no more effective (and often less effective) in WDM than in CDM. This is in contrast to Robles et al.~(\citeyear{Robles:2017}), who found that SIDM dwarfs were mostly unaffected by the addition of hydrodynamics. We find a similar result to Robles et al.: 6 out of the 8 SIDM dwarfs do not have further central (inner 500 pc) density depletion with the addition of hydrodynamics. Two halos however, m10e and m10k, do have $\sim25\%$ lower densities than their DMO counterparts. While one might be quick to attribute this to increased stellar formation (as halo m10k does have the highest $z=0$ stellar mass of all 8 SIDM halos), there are three SIDM dwarfs with more stellar mass than m10e at $z=0$ which do not show increased depletion with the addition of hydrodynamics. The SIWDM2 versions of each simulated dwarf do show further reduction in dark matter density in the dwarfs with more stellar mass at $z=0$; this is not true for any of the SIWDM3 dwarfs, however. The dwarf with the highest $\mstar$ in the suite, m10k, has twice the central density in the hydrodynamics run of SIWDM3 compared to its own DMO version.

\subsection{Rotation Curves}
In Fig \ref{fig:vcirc_1x4}, we explore rotation curves for the same halos as in Fig. \ref{fig:radden_1x4}. The DMO versions are plotted in the top row and hydrodynamical simulations are plotted in the bottom row. For the hydrodynamical simulations, we also mark the circular velocity at the 3D half-light radius $\rh$, $\vh\equiv V_\mathrm{circ}(\rh)$, for each curve with a point in matching color. While the overall normalization of the DMO rotation curves changes from halo to halo, the behavior when varying the DM theory remains similar across all halos. Introducing a non-negligible free-streaming length of the DM particle (CDM to WDM2) results in a decrease of the overall rotation curve (and therefore of $\vmax$ but little to no additional lowering of the inner rotation curve relative to $\vmax$. Introducing self-interactions results in the opposite behavior: $V_\mathrm{max}^{\rm DMO}$ remains very similar but the inner rotation velocity is lowered. Both effects are present in the SIWDM runs. 

This clean behavior seen in the DMO simulations is muddied in the hydrodynamical simulations, as the behavior of the inner density profile also depends on the effectiveness of baryonic feedback at lowering the rotation curve. In halo m10d, a dwarf in which stellar feedback has been ineffective at altering the inner density profile, the various curves are nearly identical to what is found above in the top row of DMO simulations. Halo m10d is alone in this quality, however, as all 3 of the higher $\vmax$ dwarfs depicted in the bottom row of Fig.~\ref{fig:vcirc_1x4} show significant overlap in their various DM rotation curves. Halo m10e's rotation curves for all but its WDM2 version are virtually indistinguishable within the inner 3 kpc. Looking at the rightmost column, we note that the SIWDM3 version of m10k has a nearly identical rotation curve to those found in the CDM or WDM2 versions of the dwarf. The curves by themselves are difficult to disentangle in the hydrodynamical versions for most of the simulated dwarfs, but by also considering $\rh$ and $\vh$ of each dwarf, clear differences emerge particularly for the dwarfs with lower values of $\mstar$.  While the inner rotation curves of the hydrodynamical simulations of halo m10e are nearly identical, they can be associated with a dwarf with a size of $\rh \sim300$ pc and $\vh\sim10$ km/s (as in the SIWDM2 version) or a markedly larger ($\rh \sim700$ pc, $\vh\sim20$ km/s) dwarf (as in the SIDM version). Note that the halos that have the lowest $\vmax$ and $\mstar$ in CDM (m10d and m10e) have the largest spread in $\rh$ and $\vh$. Before exploring the implications of this for observed dwarfs, it is useful to focus on how the rotation curves in each DM theory compare to one another and how they are affected by addition of hydrodynamics.

Fig.~\ref{fig:vcirc_ratio_3x4} provides a deeper look at the effects of each version of DM on the simulated dwarfs' rotation curves. Each column is dedicated to a separate (non-CDM) theory of DM; all curves are colored according to $\mstar(z=0)$ in the hydrodynamical version of each run. The first row displays, for each dwarf in each model, the ratio of the DMO rotation curve to the CDM DMO version; this shows how the physics of DM affects the dwarfs' rotation curves. WDM2 has a very similar variation with radius as its CDM counterpart, albeit shifted to a smaller magnitude (roughly 80$\%$ of the CDM value) at each radius. SIDM deviates heavily in the inner $\sim 1$~kpc from its CDM counterpart ($\sim60\%$ smaller velocity at the innermost resolved radius) but ultimately converges at large radii. Both variants of SIWDM combine the effects of WDM and SIDM, with the colder of the two (SIWDM3) showing less of an overall shift to lower rotation amplitudes than the SIWDM2 version. 

The second row of Fig.~\ref{fig:vcirc_ratio_3x4} focuses on the hydrodynamical versions of each simulation. When hydrodynamics are introduced into the simulations, the difference between CDM and alternate theories of DM becomes less clean. Three WDM dwarfs now have a higher circular velocity than their CDM counterparts in the inner $1$~kpc as opposed to a uniform shift to lower $V_{\rm circ}$. Although the SIDM simulations still have much lower values $\vcirc$ in their inner $1$~kpc compared to their CDM counterparts, the difference is reduced relative to the DMO comparison in $75\%$ of the halos. The SIWDM simulations, too, show smaller reductions than in the DMO simulations, to the point where one of the SIWDM2 dwarfs (m10m) is actually more dense than its CDM counterpart in the inner $\sim700$ pc; another of the SIWDM3 dwarfs (m10f) shows an almost one-to-one ratio with its CDM counterpart in full physics simulation. These ratios are the end result of many complex processes, so the absence of a simple correlation between dark matter density and $\mstar(z=0)$ is not surprising.

To better examine how these ratios change when moving from DMO to hydrodynamic runs, the third row of Fig.~\ref{fig:vcirc_ratio_3x4} shows the ratio of each dwarf with hydrodynamics to their DMO counterparts in each dark matter variant. The CDM ratio is plotted as a shaded region in each panel. The inclusion of hydrodynamics in CDM significantly lowers $\vcirc$ in the inner kpc when compared to the DMO simulations. However, the same is not true for the other versions of DM: in these cases, including baryonic feedback does not lower the density in the central region of each dwarf beyond what is already seen in the DMO simulations. Focusing on the simulations with self-interactions (columns 2-4), even the opposite can occur: the majority of halos in the SIDM hydrodynamical runs are more dense in the inner 1 kpc relative to their DMO versions. While this result is expected based on simulations that model SIDM as an isothermal gas in hydrostatic equilibrium within the total gravitational potential provided by dark matter and baryons \citep{Kaplinghat:2014, sameie2018}, it is reassuring to see the same effect in simulations run with actual hydrodynamics. Furthermore, this enhanced central density is present in the SIWDM simulations as well and is more prominent for larger free-streaming lengths.

\section{Discussion}
\label{sec:discussion}
Given the number of similar outcomes that DM physics plus hydrodynamics have on the central regions of dwarf galaxies, the natural question is which model (if any) compares well with observations? Modifications to CDM have been invoked to explain many discrepancies with observations of dwarf galaxies, but the successes of recent simulations in matching various observables raises the question of whether any of the discrepancies require an explanation beyond the effects of baryonic physics in CDM. We therefore compare rotation curves from the CDM simulations with data from nearby dwarf galaxies in Fig.~\ref{fig:vcirc_cdm}; the top panel shows DMO simulations while the bottom panel shows hydrodynamic simulations. Cyan points represent observational $\vh$ data of Local Field dwarfs ($10^5<\mstar<10^8\,\msun$) and are again taken from \citealt{Garrison-Kimmel:2018} and references therein (same as those plotted in Fig. \ref{fig:trip}). For purely dispersion-supported galaxies, $\vh$ is calculated using the Wolf et al.~(\citeyear{Wolf:2010}) formula relating the mass contained within the 3D half-mass radius, $\rh$, with the line-of-sight velocity dispersion $\sqrt{\langle\sigma_\mathrm{los}^2\rangle}$: 
\begin{equation}
\label{eq:wolf}
M^\mathrm{ideal}_\mathrm{Wolf}(<\rh)=\frac{3\,\rh\,\langle\sigma_\mathrm{los}^2\rangle}{G}\,.
\end{equation}
This estimator has been found to reliably recover the total enclosed mass for simulated dispersion-supported dwarf galaxies in the $\mhalo\sim10^{10}\,\msun$ range \citep{campbell2017,Gonzalez-Samaniego:2017,errani2018}. Tucana, WLM, and Pegasus also display evidence of rotational support (see \citealt{Fraternali:2009}, \citealt{Leaman:2012} and \citealt{Kirby:2014} respectively), which could cause the Wolf estimator to underestimate the true halo mass. For these three galaxies, we adopt the modified $\vh$ values presented in \citet{Garrison-Kimmel:2014}.

CDM halos in the DMO simulations have difficultly matching 6 of the dwarfs in Fig.~\ref{fig:vcirc_cdm}. 4 observed dwarfs (And XXVIII, Leo A, Cetus and IC 1613) have \textit{lower} values of $\vh$ than any of the simulations, while Tucana and NGC 6822 both have significantly \textit{higher} $\vh$ than is found in the CDM DMO runs. Though the suite was not calibrated in any way to reproduce the Local Field population of dwarfs (whether in terms of its $\vmax$ function or its environment), the addition of hydrodynamics improves the match between simulated CDM dwarfs and the low-density field galaxies around the Local Group, with only one observed point remaining inconsistent with the simulations. The dense galaxies Tucana and NGC 6822 remain without a simulation match, however, and present a greater challenge to reproduce in \fire-2 simulations, regardless of the underlying DM theory, as we discuss below. 

The results of Fig.~\ref{fig:vcirc_cdm} appear to indicate that the simulated field dwarf galaxies, all having $\mvir(z=0)\sim10^{10}\,\msun$, can reproduce the wide array of $\vh$ values measured in the Local Field at roughly the same stellar masses ($10^5<\mstar<10^8\,\msun$). While this agreement is encouraging, it does not provide the full picture: the simulated galaxies must have the correct sizes ($\rh$) while matching the stellar masses and circular velocities of observed galaxies. In Fig. \ref{fig:vcirc_2x4}, we plot the rotation curves of the dwarfs, including all DM variants, split according to $\rh$ (increasing from left to right). The top row shows DMO versions of the simulated dwarfs, while the bottom row shows those same dwarfs in the runs with hydrodynamics; the size of each dwarf in the DMO simulations is taken from the corresponding dwarf run with hydrodynamics. The cyan points with error bars again correspond to Local Field dwarf galaxies. 

The left-most panels correspond to the smallest size bin ($0<\rh<500$~pc) and contain 3 simulated CDM dwarfs and 8 observed dwarfs. Despite this, the 3 CDM halos' $\vcirc$ curves fall within the $1\,\sigma$ ($1.5\,\sigma$) error of 5 (7) of 8 halos. The 4 WDM dwarfs do an equally good job at matching the observed points. While the 2 SIDM/SIWDM2 dwarfs each provide a better fit the lower $\vcirc$ points, this comes at the cost of the small $\rh$, high $\vcirc$ points that both are too dense to be described by the SIDM/SIWDM simulations. The only major outlier is Tucana, which none of the DM models fit; the agreement is even worse in all of the WDM and/or SIDM models. While CDM is actually the best-fit model here from a $\chi^2$ perspective, the differences between CDM and the other models are not very significant if one does not consider Tucana. In the $500<\rh<750$~pc bin, the only observed point is NGC 6822, which -- similar to Tucana in the previous panels -- is a $>2\,\sigma$ outlier for all the DM models considered. At $750<\rh<1000$, there are 2 observed points. Pegasus agrees well with the CDM rotation curves while Cetus, the lower point, is a $>2\,\sigma$ outlier for CDM + WDM. While SIDM/SIWDM2/SIWDM3 do appear to provide a better fit for both points in this bin, even the lowest curves only fall within $\sim\!1.5\sigma$ of the lower point, Cetus. The rightmost panel, covering $1000<\rh<2100$ pc, shows a very similar result: the CDM version of the simulated dwarf agrees within <10$\%$ with one of the observed points (WLM) while the SIDM/SIWDM2 curves only provide marginally better matches for the lower point, IC 1613. It is worth noting that there are two observed dwarfs in this bin, but only one simulated dwarf.

To further examine this comparison and provide a more quantitative comparison, Fig.~\ref{fig:vhalf_vs_rhalf} compares observations and simulations in the $\vh-\rh$ plane. Instead of measuring $\vh=V_\mathrm{circ}(\rh)$ directly from the mass profile in the simulations, we compute $\vh$ in a similar fashion to how it is calculated for observed dwarf galaxies in order to make as fair a comparison as possible: for each dwarf, we compute the dynamical mass within $\rh$ from the stellar-mass-weighted velocity dispersion, $\sigma_\mathrm{los}$, measured within $4\,\rh$, using Eq. \ref{eq:wolf}. Each point marks the median value of $\vh$ computed over 1000 random line-of-sight projections distributed uniformly on the unit sphere (see \citealt{Gonzalez-Samaniego:2017}); error bars mark the 16th and 84th percentiles in the distributions. 

While our simulations are all of $10^{10}\,\msun$ halos, there is no reason to believe that every observed local field dwarf should reside in halos of precisely this mass. To understand how varying the halo mass might fill this parameter space, we have included a number of CDM FIRE dwarfs with halo masses between $10^{10}$ and $10^{11}\,\msun$ from Graus et al.~(in prep.)\footnote{These halos were run at a lower resolution comparable to the `low' resolution present in \citep{Fitts:2017}} as well as analytic fits based on the coreNFW profile\footnote{The coreNFW profile behaves like an NFW profile at large radii and has a core on small scales. This profile is fully described by the concentration of the NFW profile, $c$, along with two additional free parameters characterizing the inner region: the core radius, $r_c$, and the degree to which the inner profile is a core, $n$ (with $n=0$ giving no core and $n=1$ giving a completely flat core). We adopt the mean concentration at each halo mass derived from the mass-concentration relation in \citet{Dutton:2014}, fix $r_c=1.75\,\rh$ (following \citealt{Read:2016a}), and set $n=1$.} \citep{Read:2016a} for a range of halo masses. 

The coreNFW prediction for $\mvir=10^{10}\,\msun$ is shown in black and includes $1\,\sigma$ scatter of $\Delta\log_{10}(c_{200})=0.1$. We also include an estimate of the relation for ultra-faint dwarfs in green (assuming $\mvir=10^{9.5}\,\msun$). Each individual coreNFW line (and associated shaded region) is plotted along a range of $\rh$ values that is consistent with the $\mhalo-\rh$ relation of the CDM simulations. At $10^{10}\,\msun$, the predicted relation is consistent with nearly all the simulated dwarfs, regardless of the underlying DM model; all of the simulated dwarf galaxies follow essentially the same $\vh-\rh$ relation. Even the WDM2/SIWDM2 dwarfs that are physically smaller than any $10^{10}\,\msun$ CDM or SIDM dwarf still agree with expectations for 10$^{9.5}\,\msun$ dwarfs in CDM. The negligible differences between various DM models therefore mean that it is incredibly difficult to isolate any individual theory in this parameter space. The only way to break this degeneracy between DM models at these masses would either require a larger sample of observed dwarf galaxies with smaller error bars and systematic errors of $<10\%$ (a difficult prospect) or obtaining spatially-resolved rotation curves for $<400$ pc (within the half-light radii) in small dwarfs, where Fig. \ref{fig:vcirc_2x4} shows that CDM/WDM and SIDM/SIWDM could be differentiated.

While the differences between DM models in Fig. \ref{fig:vhalf_vs_rhalf} are marginal, the diversity in the observed population does appear greater than what we find in the suite of simulated dwarfs. Two field dwarfs, Cetus and IC 1613, have derived $\vh$ values that are lower than the 68$\%$ confidence region of the analytic fit.  While the low $\vh$ values would be more consistent with a lower halo mass, the large \textit{sizes} of these dwarfs are not expected for a lower halo mass given our simulations' $\mhalo-\rh$ relation. \citet{Read:2016b} have noted that the gas morphology of IC 1613 implies a state of disequilibrium, which may explain why it is not consistent with our expected relation. A $\sim30\%$ systematic difference in $\vcirc$ (e.g. from the observed $\vcirc$ under-estimating the true $\vcirc$; \citealt{Oman:2017,Verbeke:2017}) would explain the difference. We note, however, that we are plotting observationally-determined $\vcirc$ values that are derived from the galaxy kinematics, not from the underlying gravitational potential, for the simulations as well. Moreover, there are other \fire-2 CDM simulations that agree fairly well with the two points with low $\vh$ values. \citet{Chan:2018}'s sample of ultra-diffuse dwarfs ($\mvir\sim10^{11}\,\msun$) occupy the same space as Cetus and IC 1613, with $\rh\sim2$ kpc and $\vh\sim20$ km/s, and the suite of ultra-faint dwarfs in Wheeler et al.~(in prep.) contains a number of dwarfs that are directly in line with the analytic fits for halos with $\mvir\sim10^9\,\msun$.

The two `compact' dwarfs with high $\vh$ values and small half-ligh radii, Tucana and NGC 6822, are also difficult to explain with the results of our suite. Both reside much further outside the $68\%$ confidence region than the dwarfs with low $\vh$ values. Though our suite of simulations is not part of a Local-Group-like environment, \fire-2 \lcdm\ simulations that simulate Local Group analogs have found similar difficulty matching high $\vh$ values for isolated dwarf galaxies: \citet{Garrison-Kimmel:2018} found no counterparts in the ELVIS on FIRE suite to observed `compact' dwarfs with high $\vh$ values and small $\rh$ values (e.g.Tucana, NGC 6822 and even satellites such as NGC 205, NGC 147 and IC 10). It has been argued that Tucana has had a previous passage through the MW or M31 disk \citep{Teyssier:2012}, which could invalidate the assumptions of \citet{Wolf:2010} used to derive its $\vh$. 

Finally, it is worth noting that the role of ``chaos'' -- more precisely, sensitivity of high-level results to very small numerical changes -- could be significant when studying the combined effects of dark matter physics and hydrodynamics. This issue has not been studied extensively, but recent work by \citet{keller2019} and \citet{genel2019} indicates that minute changes in initial conditions can lead to macroscopic changes in the final properties of galaxies in hydrodynamical simulations. Quantifying this effect in the presence of dark matter self-interactions will be an important future step to ensure that any conclusions are fully robust to numerical effects.

\section{Conclusions}
\label{sec:conclusions}
We presented a study of isolated dwarf galaxies, comparing \lcdm~alongside WDM and SIDM (as well as SIWDM). Our simulations are cosmological and are run both with and without the hydrodynamical \fire$-2$ galaxy formation model to provide an in-depth look at how baryonic feedback interacts with different underlying DM theories. Our suite focuses on dwarfs at the $\mvir(z=0)\approx10^{10}\,\msun$ mass scale, which is relevant for the small-scale issues of \lcdm~that may be resolved through either baryonic feedback and/or alternative dark matter solutions. 

We initially looked at the global properties of isolated dwarfs simulated in different dark matter models. The simulated dwarfs have similar stellar half-mass radii, stellar velocity dispersions, and dynamical-to-stellar mass ratios to dwarfs found in the Local Field. 
In no case do we find  $\sigma_{\star}/\vcirc \approx 1/\sqrt{3}$, the maximal value attainable (when $\rh=r_{\rm max}$ of the dark matter halo). Instead, the simulated dwarfs have a limiting value of $\sigma_{\star}/\vcirc \approx 0.4$ (attained in the dwarfs with the largest values of $\rh$), with smaller dwarfs ($\rh<500$ pc) having smaller values of $\sigma_{\star}/\vcirc$ (Figure~\ref{fig:vel_ratio_vs_rhalf}). 

This finding, which is robust to varying dark matter physics, suggests that smaller dwarfs may reside in larger halos than might otherwise be inferred, though the strength of this result will depend on whether such a relation holds over a larger range in halo masses. Intriguingly, the $\rh-\mstar$ relation for the simulated dwarfs does not depend on the underlying DM theory. Since our CDM simulations result in cuspy dark matter profiles below $\mstar \sim 10^6\,\msun$ while the alternate dark matter models studied here have lower central densities for such dwarfs, the constancy of the $\rh-\mstar$ relation points to the circular velocity at fixed stellar mass (below $\mstar \sim 10^6\,\msun$ as a potential discriminator between CDM and alternate models.

Rotation curves in DMO simulations are altered in a straightforward manner. Increasing the free-streaming length of the DM particle (CDM to WDM2) results in a reduction in $V_{\rm circ}(r)$ within $r_{\rm max}$ but does not produce any differential (additional) reduction in the central circular velocity relative to $V_{\rm max}^{\rm DMO}$. SIDM results in the reverse effect: the halos have similar values of $V_\mathrm{max}^{\rm DMO}$ as their CDM counterparts but have lower circular velocities at their centers. Introducing baryons into the simulations erases many of the differences. The inner rotation curves of brighter dwarfs in CDM and WDM2 are more affected by baryonic feedback and hence more closely resemble their self-interacting counterparts, SIDM and SIWDM2. Simulating the self-interacting theories (SIDM, SIWDM2, and SIWDM3) with baryons actually served to \textit{increase} the inner rotation curve for the majority of the simulated dwarfs through the contraction of their central baryons. The effects of self-interacting DM and baryonic feedback on the inner density profile therefore do not add together. 

In brief, \textit{baryonic feedback can reduce the central density of a cuspy dark matter halo, but if there is already a core present in the halo, feedback will not appreciably lower the density profile further}. The inclusion of baryons into simulations of dwarf galaxies therefore generally serves to diminish differences that exist in dark-matter-only simulations of the dark matter models considered in this work. At the lowest $\mstar$ values simulated here, rotation curves on small scales ($r \la \rh$) provide a potential path forward for differentiating among DM models in low-$\mstar$ galaxies.  Moving forward, it will also be important to ensure that chaotic effects in numerical simulations (e.g., \citealt{keller2019, genel2019}) including baryonic physics and non-standard dark matter physics are minimal or are at least well understood.

In order to comprehensively address the small-scale issues in field galaxies, simulated dwarfs must simultaneously match the measured $\vh$ and $\rh$ values of Local Field dwarfs. Performing mock observations on the suite of simulated dwarfs, we find that all of our DM models follow a similar median $\vh-\rh$ relation, with no evidence of systematic differences between any of the models. All of the models therefore do comparably well at fitting the observational data. While a small number of observed dwarfs are outliers (both above and below the simulation-derived relation presented here), this may be related to the sample size. Larger samples of \fire-2 CDM simulations are able to routinely reproduce these outlier observations, with the possible (interesting) exception of Tucana-like dense and compact dwarfs. A larger sample of simulated dwarfs at this mass scale with enough resolution to reliably determine $\rh$ is needed to better understand the level of scatter expected in the $\vh-\rh$ relation and whether or not dwarfs like Tucana fall within it. It is also important to investigate how halo mass affects dwarfs in alternative models of DM;  further simulations are needed to understand whether modifications of CDM can explain the full range of observed dwarf galaxy properties.

\section*{Acknowledgments}
AF thanks Bonnie and Emily Collins for the valuable discussions, and Alexander Knebe and Oliver Hahn for making \ahf\ and \music, respectively, publicly available. MBK and AF acknowledge support from the National Science Foundation (grant AST-1517226). MBK was also partially supported by NSF CAREER grant AST-1752913 and NASA through grant NNX17AG29G and HST grants AR-12836, AR-13888, AR-13896, AR-14282, AR-14554, GO-12914, and GO-14191 awarded by the Space Telescope Science Institute (STScI), which is operated by the Association of Universities for Research in Astronomy (AURA), Inc., under NASA contract NAS5-26555. JSB was supported by NSF AST-1518291, HST-AR-14282, and HST-AR-13888. Support for PFH was provided by an Alfred P. Sloan Research Fellowship, NASA ATP Grant NNX14AH35G, and NSF Collaborative Research Grant \#1411920 and CAREER grant \#1455342. CAFG was supported by NSF through grants AST-1517491, AST-1715216, and CAREER award AST-1652522, by NASA through grants NNX15AB22G and 17-ATP17-0067, and by a Cottrell Scholar Award from the Research Corporation for Science Advancement. DK was supported by NSF grant AST-1715101 and the Cottrell Scholar Award from the Research Corporation for Science Advancement. AW was supported by NASA through ATP grant 80NSSC18K1097 and grants HST-GO-14734 and HST-AR-15057 from STScI. This work used computational resources of the University of Texas at Austin and the Texas Advanced Computing Center (TACC; \url{http://www.tacc.utexas.edu}), the NASA Advanced Supercomputing (NAS) Division and the NASA Center for Climate Simulation (NCCS) through allocations SMD-15-5902, SMD-15-5904, SMD-16-7043, and SMD-16-6991, and the Extreme Science and Engineering Discovery Environment (XSEDE, via allocations TG-AST110035, TG-AST130039, and TG-AST140080), which is supported by National Science Foundation grant number OCI-1053575. 
\label{lastpage}
\bibliography{main}
\appendix
\section{Impact of Resolution}
\label{sec:appendixa}
\begin{figure*}
\includegraphics[width=\textwidth]{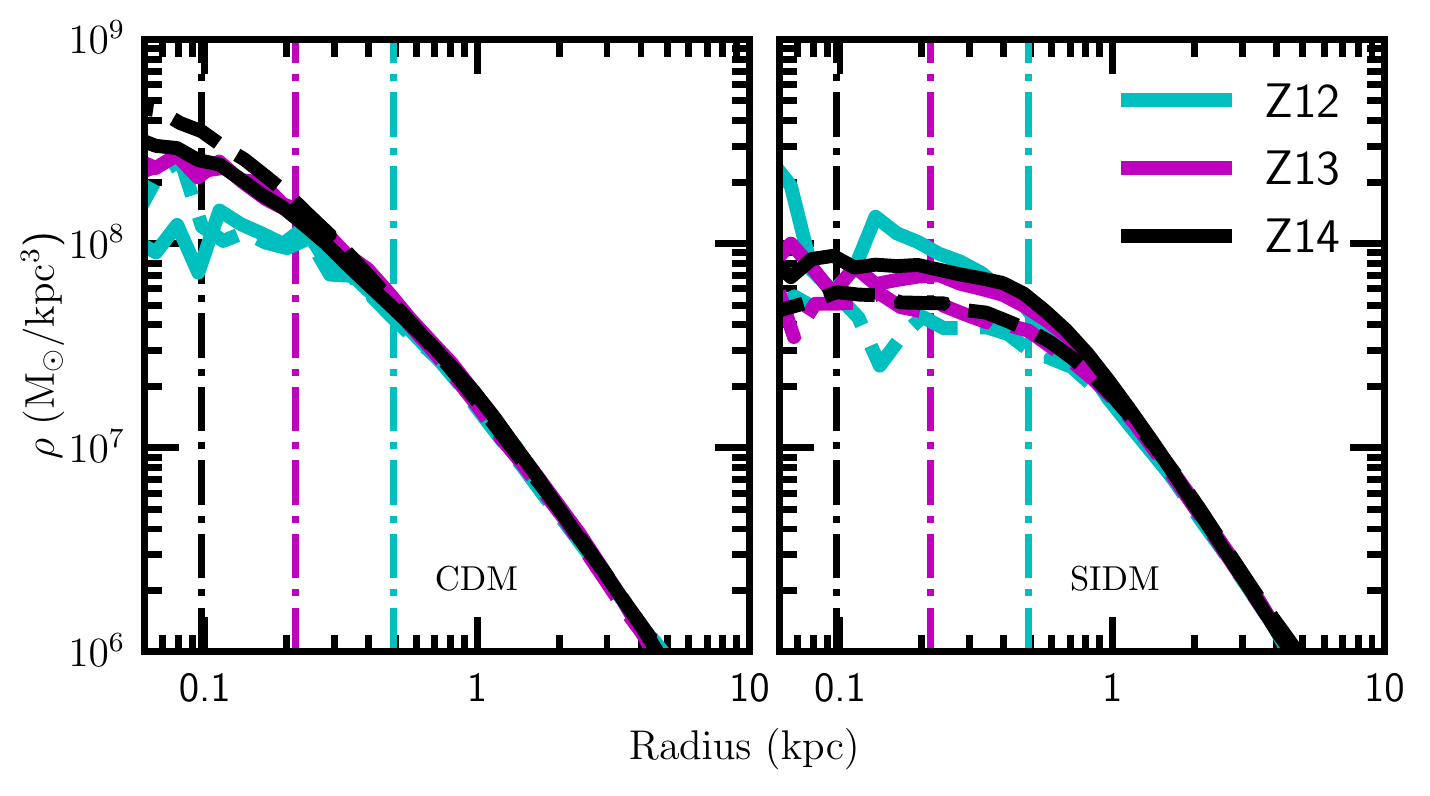}
    \vspace{-0.5cm}
  \caption{Radial density profile convergence of halo m10b in both CDM and SIDM. Each panel shows the density profile for hydrodynamical (solid) and DMO (dashed) runs of an individual halo at three resolutions: Z12 (cyan), Z13 (magenta), and Z14 (black). The Power radius for each run is marked by a vertical dotted line of the corresponding color and provides a relatively conservative approximation for where each density profile deviates from its higher resolution counterpart (i.e., density profiles are essentially perfectly converged for $r \ge \rpower$ and are converged to better than $\sim 20\%$ in density for $r \gtrsim 0.5 \,\rpower$). At our fiducial resolution (Z13),  $\rpower$ is $\approx 200\,{\mathrm{pc}}$ for the DMO simulations.}
\label{fig:res_rd_comp}
\end{figure*}

To understand the convergence properties of our simulations, we look in particular at halo m10b in both CDM and SIDM. We have run halo m10b at 2 times poorer (better) force and 8 times poorer (better) mass resolution. Our lowest resolution is referenced as Z12, our fiducial as Z13 and our highest as Z14. In Fig. \ref{fig:res_rd_comp}, we present the convergence of radial density profiles of halo m10b at the three different resolution levels in both hydrodynamical (solid) and DMO (dashed) versions of the simulation. \citet{Power:2003} proposed that an estimate of numerical convergence radius for density profiles in dark matter simulations is the radius where the two-body relaxation time exceeds 60\% of the current age of the Universe (corresponding to the radius enclosing $\sim2500$ particles); Fig.~\ref{fig:res_rd_comp} demonstrates that this Power criterion provides a conservative measure of numerical convergence. We refer to this ``Power radius'' (calculated just from dark matter particles) as our reference ``convergence radius'' throughout (and note that $\sim20\%$ convergence in density can be obtained at radii enclosing just $\sim200$ particles). The Power radius for each hydro simulation is marked with a dotted line, with color matching the corresponding density profile, in the figure. In each case, the density profiles agree well between the two resolutions for all converged radii.

\begin{figure}
\includegraphics[width=\columnwidth]{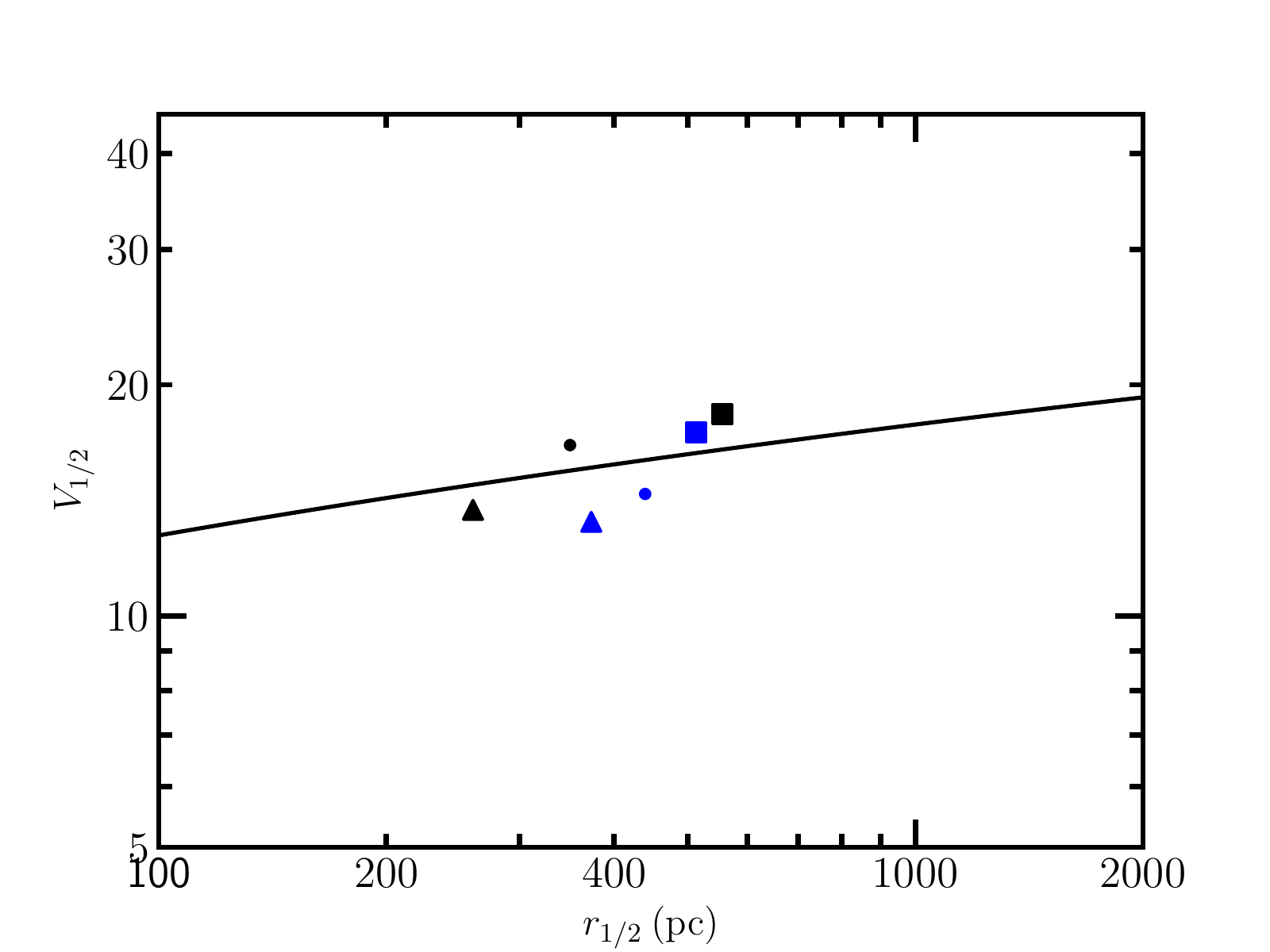}
    \vspace{-0.5cm}
  \caption{$\vh$ vs $\rh$. CDM in black, SIDM in blue. Low resolution is represented by squares, fiducial resolution by circles and the high resolution by triangles. The black line is the best fit line for all CDM simulations (including those from Graus et al., in preparation). Increased resolution leads to smaller $\rh$ and a lower $\vh$. The decrease in $\vh$ is mostly due to the smaller extent of $\rh$, as Fig.~\ref{fig:res_rd_comp} shows that the changes in the inner density profile with increased resolution are not enough to account for the change in $\vh$. While increased resolution does move our points substantially, it is encouraging that they move along the best fit line of the CDM simulations, leaving our main conclusions unchanged.}
\label{fig:res_comp}
\end{figure}

As was noted in \citet{Fitts:2018}, though much of galaxy properties of halo m10b appeared to be converged across resolution levels, the 3D stellar half-mass radius decreased in the CDM versions of halo m10b as the resolution was increased. To examine how this may affect our $\vh-\rh$ relation, we plot the three resolution levels of the CDM and SIDM simulations for halo m10b in Fig. \ref{fig:res_comp}. The CDM simulations show a decrease of 53$\%$ in $\rh$ when going from Z12 to Z14 while the SIDM simulations see only 25$\%$ of a reduction in size across our range of resolution. This reduction in $\rh$ is accompanied by a reduction in $\vh$, as expected. Given the strong convergence in density profiles seen in Fig. \ref{fig:res_rd_comp}, a smaller $\rh$ corresponds to a smaller amount of mass contained within that radius. Hence while the Z14 runs show lower $\vh$ values than the corresponding Z12 runs, this is always matched with a corresponding decrease in $\rh$. Looking at \ref{fig:res_comp}, increased resolution only translates to moving along the best fit line of the CDM simulations and does not appear to affect our main results.

\end{document}